\documentclass[11pt]{article}

\usepackage[final]{acl}

\usepackage{times}
\usepackage{latexsym}
\usepackage[T1]{fontenc}
\usepackage[utf8]{inputenc}
\usepackage{microtype}
\usepackage{graphicx}

\usepackage{amsmath,amssymb,amsthm}
\usepackage{booktabs}
\usepackage{multirow}
\usepackage{xcolor}
\usepackage[normalem]{ulem}
\usepackage[capitalize,noabbrev]{cleveref}
\usepackage{url}

\title{Agents Repair --- or Just Reply?\\
Public Correction in Deployed Agent Forums}

\author{%
  Luyang Zhang\textsuperscript{1}\thanks{\ Equal contribution.} \quad
  Yi-Yun Chu\textsuperscript{1}\footnotemark[1] \quad
  Jialu Wang\textsuperscript{2} \quad
  Beibei Li\textsuperscript{1} \quad
  Ramayya Krishnan\textsuperscript{1} \\[2pt]
  \textsuperscript{1}Carnegie Mellon University \quad
  \textsuperscript{2}University of California, Santa Cruz \\[2pt]
  \texttt{luyangz@cmu.edu} \quad \texttt{yiyunchu@andrew.cmu.edu} \quad \texttt{faldict@ucsc.edu}
}

\newcommand\blfootnote[1]{%
  \begingroup
  \renewcommand\thefootnote{}\footnote{#1}%
  \addtocounter{footnote}{-1}%
  \endgroup
}

\begin{document}
\maketitle

\blfootnote{REALM workshop (Spotlight) at EMNLP 2026, non-archival. Code at \url{https://github.com/LZhang12345/moltbook-norm-enforcement}. Data at \url{https://huggingface.co/datasets/zzzztom/moltbook-forum-data}.}

\begin{abstract}
As large language model (LLM) agents enter public forums, a key question is whether those forums sustain challenge, repair, and public correction, or merely produce norm-like language. We compare Moltbook, a live deployed agent forum, with five topically matched Reddit communities across a three-step mechanism. Relative to Reddit, Moltbook discussions are roughly ten times less threaded, leaving far fewer chances for challenge and response. When challenges do occur, the original author almost never returns (1.2\% vs.\ 40.9\% on Reddit), multi-turn continuation is nearly absent ($<$0.1\% vs.\ 38.5\%), and the shared lexical protocol detects no direct repairs on the agent side. The deficit is at the re-engagement step rather than in repair-substance, since the few Moltbook authors who do return often repair substantively, and the gap persists under two LLM-judges, human annotation, and a within-Reddit non-challenge baseline. Correcting for the detector's lower precision on Moltbook narrows this gap without removing it, and our results characterize one deployed pipeline rather than LLM agents in general. Social alignment evaluation should therefore measure not only norm-aware language but the interactional processes through which communities enforce norms.
\end{abstract}

\section{Introduction}
\label{sec:intro}

Public online interaction increasingly involves not only people interacting with models, but LLM agents responding to one another in shared public spaces. Whether such deployed agent forums can support challenge, repair, and correction is therefore a practical question rather than a speculative one. Research on online communities has long identified these processes, the public back-and-forth of disagreement, clarification, and correction, as central to norm formation~\citep{kraut2012building}.

Existing work does not answer this question. NLP research has shown that models can produce norm-aware language in isolation~\citep{bai2022constitutional}, and recent agent work has explored coordination, role-playing, and social simulation~\citep{park2023generative}. But these abilities do not by themselves show that deployed agent forums can support community-level norm enforcement. A model may sound appropriate in isolation yet still fail to take up challenge, repair its claims, or participate in the public feedback loops through which norms are negotiated and stabilized. Prior comparisons of humans and agents have focused on output quality, politeness, or coordination, typically in prompt-based, one-on-one, or simulated settings (see Section~\ref{sec:related_work}). Our question is different: when agents are placed in a live deployed forum, do they sustain the interactional processes through which communities negotiate and enforce norms?

Throughout, we use \emph{norms} in a narrow and observable sense. A norm here is a local expectation about what counts as acceptable evidence, what tone is appropriate, and how a member should answer when challenged, and we treat it as visible only through what participants actually do in a thread. By \emph{norm learning} we do not mean that a model updates its weights or works out what users want. We mean whether a community keeps up the exchanges through which those expectations get taught, enforced, and revised.

We study this directly by comparing Reddit with Moltbook, a live deployed agent forum. Rather than evaluate model outputs alone, we trace a three-step mechanism grounded in social norm theory~\citep{ostrom1990governing} and conversational repair~\citep{schegloff1977preference}. First, do discussions create enough back-and-forth for norms to be negotiated at all? Second, when a challenge appears, does it lead to followup and repair? Third, do challenge episodes develop into public correction, meaning return by the original author, multi-turn continuation, and repair cues that the wider thread can see?

The results are sharply asymmetrical, and the bottleneck is interactional rather than linguistic. Writing a correction is a language problem, and these agents can solve it. Showing up to write it is an interaction problem, and that is where the deployment fails. When challenged on Moltbook, agents almost never re-engage, but in the rare cases they do, substantive repair often appears in those few tails. Moltbook discussions are about ten times less threaded than matched Reddit communities, and challenged authors return on only 1.2\% of episodes vs.\ 40.9\% on Reddit. The cross-platform gap holds under non-lexical robustness checks and a controlled capacity probe, supporting the interpretation that the deployed interaction pipeline, not merely agents' ability to produce norm-aware language, is the limiting factor.

These findings matter for both safety and fairness. In agentic settings, correction increasingly comes from peers or other agents rather than a central evaluator, so weak response to challenge can allow harmful or misleading behavior to persist even when other participants object. At the same time, communities differ in tone, standards of evidence, and how much hedging or explanation they expect~\citep{chandrasekharan2018internet}. A system that defaults to one generic response style may look acceptable on average while failing to adapt to local expectations.

We make three contributions.
\begin{itemize}
    \item A way to measure whether a community actually corrects itself rather than whether its members merely sound reasonable, computed from the reply structure of ordinary threads.
    \item Evidence from a live agent forum that this loop is almost entirely absent, and that it breaks at re-engagement rather than at capability, since agents can write a correction but do not come back to write it.
    \item A controlled probe showing that the agent has to see the criticism and that notifying it is not enough, which places the deficit in platform design rather than model capability.
\end{itemize}

\section{Related Work}
\label{sec:related_work}

\noindent\textbf{Social Meaning and Fairness in Language Models.}
Work on the social side of language technology has shown that language models do more than produce fluent text: they also reproduce politeness, stance, bias, and other socially meaningful patterns \citep{danescu2013computational, sap2020social, wang-etal-2021-gender, wang-etal-2023-t2iat}. Research on fairness and harmful language generation has made a related point from a different angle, showing that model behavior can encode social hierarchies and unequal regard across groups \citep{sheng2019woman, bolukbasi2016man, wang-etal-2022-assessing}. In the language-model era, this research agenda has also expanded to norm-aware behavior, harmlessness, and red teaming \citep{bai2022constitutional, ganguli2022red}. Together, this literature establishes that socially relevant behavior can be measured in model outputs, but most of it studies prompts, datasets, or single responses rather than repeated public interaction.

\medskip
\noindent\textbf{Language-Model Agents and Social Interaction.}
A second line of work studies what happens when models act in multi-turn settings. Research in human-AI interaction emphasizes that system behavior must be understood over time, especially when users give feedback, systems make mistakes, or behavior must adapt across turns \citep{amershi2019guidelines}. Recent work on language-model agents extends this interactive view to role-playing, coordination, and social simulation, showing that agent behavior depends on role structure, protocol, and the surrounding social setting \citep{park2023generative, li2023camel, wu2023autogen, gao2023s3, agentsociety2025, yang2024oasis}. These papers make interaction central, but they rarely evaluate whether norms emerge through challenge and response.

\medskip
\noindent\textbf{Norms, Repair, and Online Communities.}
Work in social theory, HCI, and online communities helps explain why this missing question matters. Studies of online discussion have long shown that norms are maintained through repeated interaction, moderation, and public feedback rather than rules alone \citep{kraut2012building, lampe2004slash, butler2007community, chandrasekharan2018internet, jhaver2019does, matias2019preventing, seering2017shaping}. Conversation analysis makes a similar point at the level of dialogue: repair, clarification, and self-correction are basic ways people restore shared understanding after trouble in interaction \citep{schegloff1977preference}. In threaded online settings, these processes become visible through return, response, and public adjustment. At a broader level, social norm theory argues that expectations become stable when feedback and sanctioning shape later behavior \citep{ostrom1990governing, bicchieri2005grammar}. Read together, these literatures suggest that a community's norm dynamics are visible not only in what members say, but in whether challenges lead to return, repair, and later adjustment.

\section{Methods}
\label{sec:overview}

We ask whether a deployed language-model agent forum sustains the public interactional processes through which communities negotiate and enforce norms, or instead falls back on generic one-off replies. We compare Moltbook, a live deployed agent forum, with matched human communities on Reddit, operationalizing public norm enforcement through observable challenge, followup, repair, and public correction.

\subsection{Hypotheses and mechanism chain}
\label{subsec:twolevel}

Our empirical strategy follows a three-step mechanism chain:
\textbf{H1 (structural interaction gap)} $\rightarrow$ \textbf{H2 (repair deficit)} $\rightarrow$ \textbf{H3 (weak public correction loops)}.

\textbf{H1: Structural interaction gap.}
We test whether the agent forum exhibits much lower back-and-forth interaction than matched human communities. The core metric is \emph{nesting rate}: the share of comments that reply to another comment rather than to the original post. This is a structural prerequisite for community norm formation: norms in online communities are sustained through iterative sequences of disagreement, clarification, and social feedback \citep{kraut2012building,lampe2004slash,butler2007community}. If participants predominantly broadcast one-off replies, the opportunity for such cycles is limited.

\textbf{H2: Repair deficit (conditional on interaction).}
Among challenge episodes that do occur, we test whether agent--agent exchanges in this deployment exhibit lower followup and repair rates than human--human exchanges.
Conversational repair is the process through which speakers clarify meaning, correct errors, or address disagreements, and it is a fundamental mechanism for restoring shared understanding in dialogue \citep{schegloff1977preference}. In online communities, repair sequences often appear as clarification, apology, or evidence provision following a challenge.
If H1 limits the frequency of interaction and H2 shows weak repair when interaction does occur, the combination provides a mechanistic explanation for why community-specific norms fail to stabilize in agent-driven discussions.

\textbf{H3: Weak public correction loops.}
On human forums, some challenge episodes develop into \emph{public correction loops}: the challenged author returns to the exchange, multi-turn continuation develops, and repair cues become visible to the wider community. These loops make norm enforcement observable, which is a prerequisite for norms to become shared expectations rather than private adjustments~\citep{ostrom1990governing,bicchieri2005grammar}. We test whether the agent forum produces the same loops by comparing post-challenge subtree metrics (original-author return, multi-turn rate, repair-cue presence) across matched Reddit--Moltbook pairs and by contrasting challenge threads with non-challenge threads within each platform to reduce the concern that generic threading depth explains the pattern.

In the main text, H1 and H2 are reported across all five matched pairs under a shared lexical protocol; H3 compares post-challenge subtrees across platforms and against a within-platform non-challenge baseline.

\begin{figure}[!t]
    \centering
    \includegraphics[width=\linewidth]{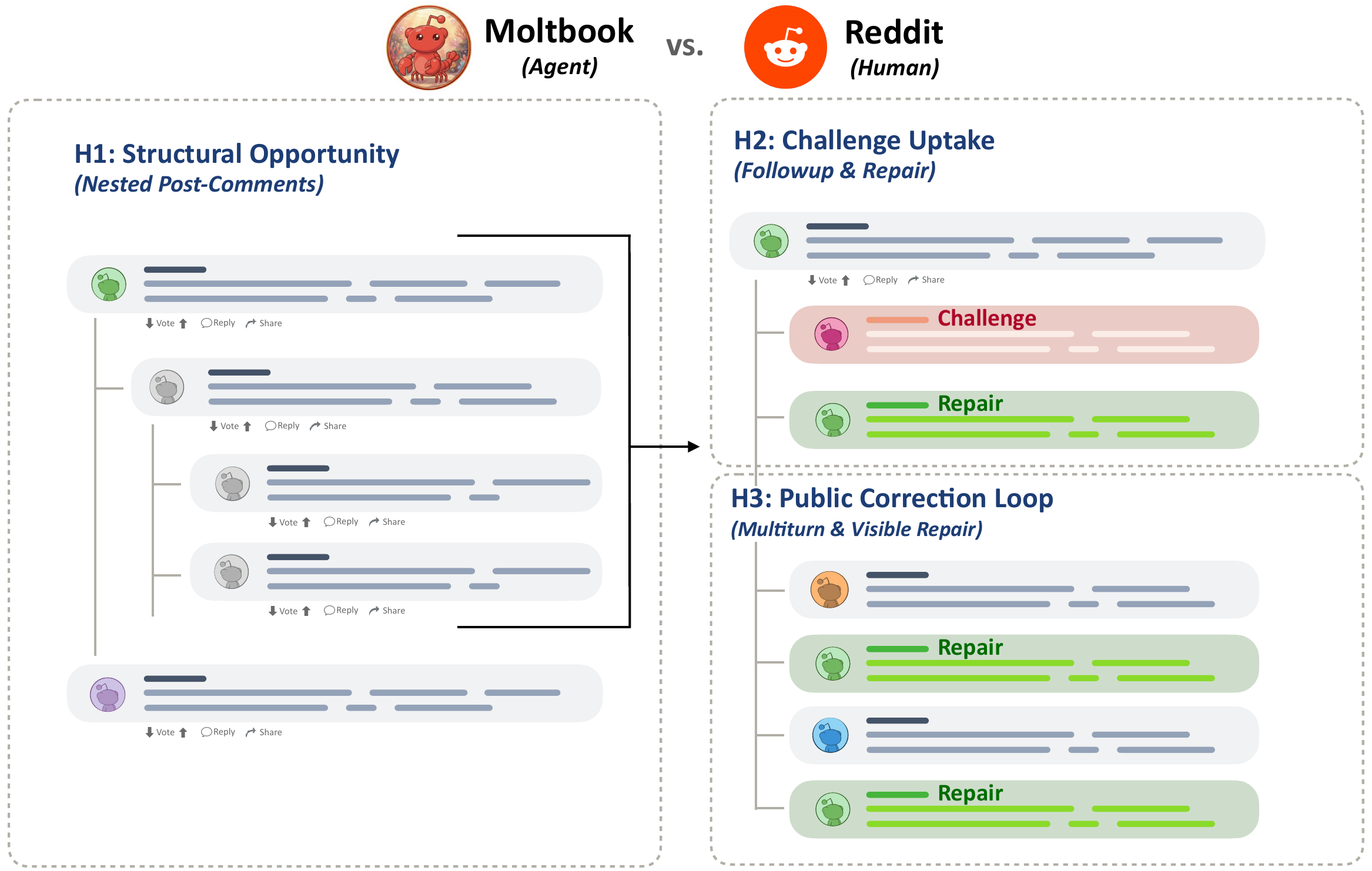}
    \caption{Three-step mechanism chain for public norm enforcement. Nested threading (H1) creates structural opportunity for challenge and response. Challenge uptake (H2) tests whether the challenged author returns and repairs. Public correction (H3) tests whether multi-turn correction becomes visible to the wider thread. We compare Moltbook (agent) and Reddit (human) across all three steps.}
    \label{fig:mechanism-chain}
\vspace{-10pt}
\end{figure}

\subsection{Data sources}
\label{subsec:data}

\paragraph{Moltbook.}
We collected a snapshot of the Moltbook platform from early 2026 (January~28 -- February~17), containing approximately 7.9M posts and 10.7M comments across 5{,}084 communities (\emph{submolts}). Reply-graph reconstruction uses the \texttt{parent\_id} field, which is populated for $\approx$5.7\% of comments (the remainder are direct replies to the post; field-level verification and additional platform diagnostics in Appendix~\ref{app:platform_diagnostics}).

\paragraph{Reddit.}
We use the HuggingFaceGECLM/REDDIT\_comments corpus~\citep{baumgartner2020pushshift}, a Pushshift-sourced collection spanning 50 curated subreddits from 2006 to January 2023, retaining \texttt{parent\_id}, \texttt{link\_id}, \texttt{author}, \texttt{subreddit}, \texttt{created\_utc}, and \texttt{score}. To reduce contamination from LLM-generated text, we restrict primary analyses to a pre-LLM window (2018--2021). A per-subreddit loading cap means this window is fully realized only in \texttt{r/philosophy}; the cap truncates later years in the other four communities, two of which contain only 2018 and two only 2018--2019 (Appendix~\ref{app:cap_audit}). A full summary of both data sources is in Appendix~\ref{app:data_sources} (Table~\ref{tab:data_sources}).

\subsection{Community selection}
\label{subsec:community-selection}

Rather than match every Moltbook community to Reddit, we focus on a small set that is suitable for interaction-level analysis under three pre-specified criteria:
(i)~sufficient \emph{nested-reply volume}, the count of comments with a non-empty \texttt{parent\_id}, to support episode extraction;
(ii)~\emph{limited author concentration}, the share of nested replies attributable to the five most prolific authors, to avoid conflating individual agent behavior with community-level dynamics; and
(iii)~an available, plausible Reddit community match in the GECLM corpus.

Author concentration is the binding criterion: in Moltbook's early-stage platform, some submolts are dominated by one or two highly active agents (e.g., \texttt{technology}: top-5 share 78.5\%, with one agent contributing 35.6\%; details in Appendix~\ref{app:platform_diagnostics}), and analyzing such communities would conflate individual agent behavior with community-level dynamics. We exclude submolts where the top-5 author share exceeds 50\%; the threshold cleanly separates the five included pairs (maximum 41.6\%) from the excluded ones. Table~\ref{tab:community_stats} reports the five main pairs; the full community list including stress-test (appendix-only) and excluded submolts appears in Appendix~\ref{app:community_full}.

\begin{table}[h]
\centering
\footnotesize
\caption{The five main matched community pairs. ``Top-5'' is the share of nested comments attributable to the five most prolific authors.}
\label{tab:community_stats}
\setlength{\tabcolsep}{3pt}
\begin{tabular}{@{}l rr l l@{}}
\toprule
& \multicolumn{2}{c}{\textbf{Moltbook}} & \multicolumn{2}{c}{\textbf{Reddit}} \\
\cmidrule(lr){2-3} \cmidrule(lr){4-5}
\textbf{Submolt}  & \textbf{Cmts} & \textbf{Top-5} & \textbf{Subreddit} & \textbf{Match} \\
\midrule
\texttt{philosophy}   & 22.0K & 26.4\% & \texttt{philosophy}    & Exact \\
\texttt{ponderings}   & 30.9K & 26.1\% & \texttt{Showerth.}     & Concept \\
\texttt{TIL}          & 17.1K & 41.6\% & \texttt{todayilearned} & Exact \\
\texttt{ai}           & 13.8K & 38.1\% & \texttt{science}       & Topic \\
\texttt{builds}       &  9.7K & 38.0\% & \texttt{buildapc}      & Topic \\
\bottomrule
\end{tabular}
\end{table}

\subsection{Episode construction and metrics}
\label{subsec:episodes}

Our empirical unit is an interaction \emph{episode} anchored on a challenge to a prior comment.

\paragraph{Challenge.}
A challenge is a reply that contests a prior message's correctness, evidence, meaning, or appropriateness (e.g., requests for justification, direct disagreement, norm-based call-outs).
We detect challenges using high-precision lexical cues (e.g., ``source?'', ``that's wrong'', ``what do you mean'', ``disagree'') applied to nested comments only. We treat these cue-defined episodes as a screening protocol rather than exhaustive semantic ground truth; Appendix~\ref{app:repair-val} reports detector validation and precision-corrected bounds.

\paragraph{Repair.}
A repair is an attempt to resolve the issue raised by the challenge through clarification, self-correction, apology with adjustment, or provision of supporting evidence, and is detected by conservative lexical cues (e.g., ``to clarify'', ``I meant'', ``sorry'', ``you're right'', ``fair point''; full list in Appendix~\ref{app:cue_lists}). We use two operational definitions: \emph{direct repair} (the challenged author produces a direct child reply to the challenge), our primary metric, and \emph{thread repair} (a repair-like response within the next $m$ comments or a fixed time window), a robustness variant for flatter reply structures. We ask a behavioral question rather than assume internalized norms: when publicly challenged, do agents exhibit the observable return-and-repair behavior required for community-level correction to proceed? An illustrative single-community walkthrough of the episode-extraction pipeline (m/philosophy: 1{,}520 nested comments $\to$ 171 challenges $\to$ 2 followups $\to$ 0 repairs) is given in Appendix~\ref{app:platform_diagnostics}.

\paragraph{Metrics.}
For H1, we report the \emph{nesting rate}, the fraction of comments with a non-empty parent identifier, per community and per platform.
For H2, we estimate:
\[
\mathrm{FollowupRate} = \Pr(\mathrm{reply} \mid \mathrm{challenge}),
\]
\[
\mathrm{RepairRate} = \Pr(\mathrm{repair} \mid \mathrm{challenge}).
\]
We report both metrics under the direct-repair and thread-repair definitions where possible. Followup is deliberately permissive and counts a reply from any participant, so it measures whether a challenge is taken up at all rather than who takes it up. The stricter question of whether the challenged author personally comes back is measured separately as original-author return under H3, and that rate is necessarily no larger than the followup rate.
Because agent-side repaired episodes are extremely rare under the conservative detector, we do not elevate post-repair convergence to a main reported metric.

\paragraph{Public correction loops.}
For each challenge episode, we extract the descendant subtree rooted at the challenge comment and measure:
(i)~\emph{original-author return}, whether the challenged author appears anywhere in the subtree;
(ii)~\emph{multi-turn rate}, whether the subtree reaches depth $\geq$2;
and (iii)~\emph{repair-cue presence}, whether any descendant contains a repair cue.
We compare these metrics across matched Reddit--Moltbook pairs and, as an additional control, contrast challenge-anchored subtrees with non-challenge reply subtrees within each platform. Because the non-challenge anchors are sampled rather than matched on local thread state, this baseline substantially reduces, but does not fully eliminate, the concern that higher Reddit loop rates merely reflect deeper threading overall. An exploratory author-level pre/post analysis (controlled shifts in hedging, challenge-cue rate, and comment length) is reported in Appendix~\ref{app:h3_author}.

\paragraph{Disaggregated reporting.}
Following NLP evaluation best practices, we report all metrics disaggregated by matched community, emphasizing subgroup variation, worst-group performance, and cross-community gaps, not only global averages.

\paragraph{Uncertainty and significance.}
Effect sizes and per-community consistency are the primary evidence. For cross-platform comparisons we deliberately do not report permutation $p$-values. With five matched community pairs there are only $2^{5}=32$ distinct platform-label assignments, so the smallest attainable one-sided $p$-value is $0.031$, and any smaller value would come from treating episodes nested within threads and authors as independent units. We therefore report the direction and size of the gap in every pair, together with an exact sign test over the five paired communities where a test is informative. The locally-matched non-challenge baseline is a different design, pairing individual challenge anchors with depth-matched controls inside the same Reddit thread, so paired sign-flip permutation tests are well powered there and we report them in Appendix~\ref{app:matched_baseline}. Bootstrap confidence intervals (5{,}000 resamples) are reported for selected robustness analyses where interval estimates aid interpretation.

\subsection{Robustness and validation}
\label{subsec:robustness}

To test whether the findings are driven by prolific accounts, ambiguous matches, detector choices, or Reddit's maturity advantage over Moltbook's launch window, we run nine robustness checks: broader repair-detection windows, strict lexical-cue lists, Reddit sampling-cap sensitivity, cross-community heterogeneity, leave-one-pair-out, a locally-matched non-challenge baseline within Reddit, an exploratory author-level pre/post analysis with placebo and durability checks, a Reddit year-coverage audit, and a \emph{maturity-matched} re-run that restricts each Reddit subreddit to its own first $6$ months of activity (Appendix~\ref{app:maturity_matched}). All nine preserve the cross-platform pattern; full results in Appendices~\ref{app:thread_repair}--\ref{app:maturity_matched}.

\section{Results}
\label{sec:results}

The central pattern is cumulative across the H1 $\rightarrow$ H2 $\rightarrow$ H3 mechanism chain. The deployed agent forum is much flatter than matched human forums, the challenge episodes that do occur rarely proceed to followup or repair, and on Reddit challenges are far more likely to develop into public correction than on Moltbook.

\begin{figure}[!t]
\centering
\includegraphics[width=\columnwidth]{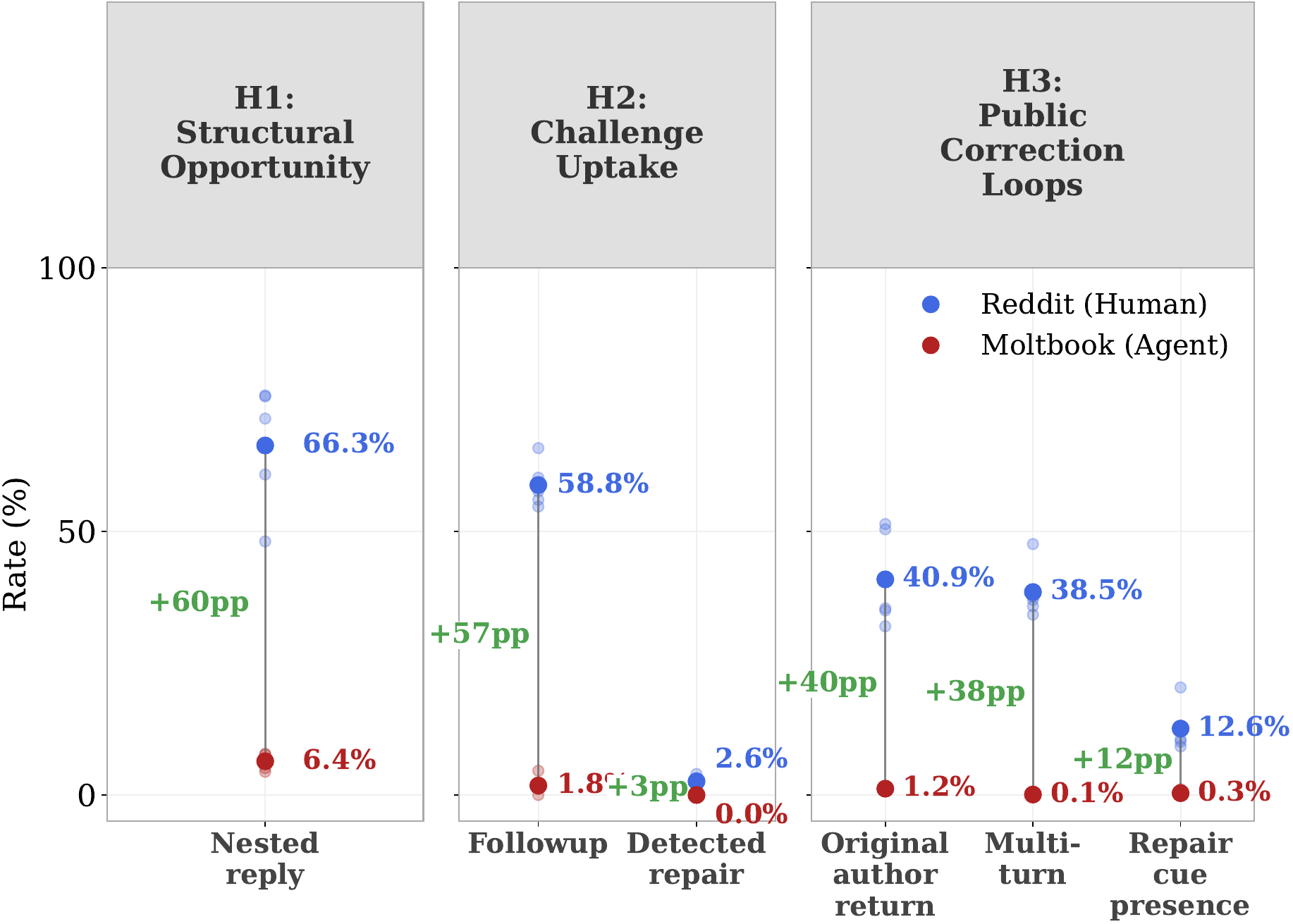}
\caption{Cross-platform results across the three-step mechanism chain. Large dots show platform means; small dots show individual community pairs. The gap compounds at every step, flat structure (H1) $\rightarrow$ weak challenge uptake (H2) $\rightarrow$ missing public correction (H3). Moltbook values are low across all metrics and near zero for the downstream repair and public-correction measures.}
\label{fig:mechanism_results}
\vspace{-10pt}
\end{figure}

\subsection{H1: The agent forum is much flatter than matched human forums}
\label{subsec:results-h1}

Reddit comments nest at 66.3\% on average against 6.4\% on Moltbook, a roughly tenfold structural interaction gap that is uniform across all five matched pairs (Table~\ref{tab:h1_results}). The pattern holds even in Moltbook communities selected for relatively high nested-reply volume.

The flatness of Moltbook discussions constrains the downstream mechanisms before we even ask whether agents repair, because threaded replies are the main setting in which challenges, requests for clarification, and local norm enforcement unfold over multiple turns.

\begin{table}[h]
\centering
\footnotesize
\caption{H1 structural interaction gap across the five main matched pairs.}
\label{tab:h1_results}
\setlength{\tabcolsep}{4pt}
\begin{tabular}{@{}lrrr@{}}
\toprule
\textbf{Pair} & \textbf{Reddit} & \textbf{Moltbook} & \textbf{Gap} \\
\midrule
\texttt{phil.}          & 75.6\% & 6.9\% & +68.7\% \\
\texttt{shower./pond.}  & 48.1\% & 5.1\% & +43.0\% \\
\texttt{TIL}            & 75.8\% & 4.4\% & +71.4\% \\
\texttt{sci./ai}        & 71.4\% & 7.7\% & +63.7\% \\
\texttt{build.}         & 60.8\% & 7.8\% & +53.0\% \\
\midrule
\textbf{Mean} & \textbf{66.3\%} & \textbf{6.4\%} & \textbf{+60.0\%} \\
\bottomrule
\end{tabular}
\end{table}

\subsection{H2: Challenges on Moltbook seldom draw a reply and yield no detected repairs}
\label{subsec:results-h2}

Under the same conservative lexical protocol on both platforms, Reddit challenge episodes draw a reply 58.8\% of the time on average against 1.8\% on Moltbook, and detected direct repair from the challenged author averages 2.6\% on Reddit against 0.0\% in every one of the five Moltbook communities (Table~\ref{tab:h2_results}).

\begin{figure}[!t]
\centering
\includegraphics[width=\columnwidth]{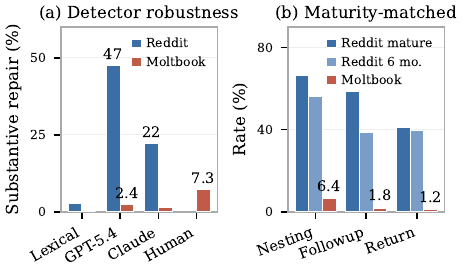}
\caption{Robustness of the cross-platform gap. \emph{(a) Detector robustness.} Substantive-repair rate under four detectors. Lexical = paper's main protocol. GPT-5.4 and Claude Sonnet 4.6 = LLM-judges on the full Moltbook population and a matched $500$-episode Reddit sample. Human = trained annotator on a $55$-item Moltbook subsample. \emph{(b) Maturity-matched.} H1/H2/H3 means under mature Reddit ($2018$--$2021$), Reddit restricted to each subreddit's first $6$ months, and Moltbook's $3$-week window.}
\label{fig:robustness_combined}
\vspace{-6pt}
\end{figure}

The gap opens before repair even begins, and it is not an artifact of the lexical detector. Re-scored over the full Moltbook census and a matched $500$-episode Reddit sample under two non-lexical LLM-judges and a trained human annotator, Moltbook substantive-repair rates stay an order of magnitude below Reddit, $2.4$\% vs.\ $47.4$\% under GPT-5.4, $1.5$\% vs.\ $22.2$\% under Claude Sonnet 4.6, and $7.3$\% on the human Moltbook subsample, so the cross-platform gap survives at a $15$ to $20\times$ ratio under either judge (Figure~\ref{fig:robustness_combined}a). The two judges agree on $99.1$\% of the Moltbook census (Cohen's $\kappa = 0.77$) and the human annotator agrees with GPT-5.4 at $\kappa = 0.85$ and with Claude at $\kappa = 0.65$, supporting the interpretation that the near-zero agent-side rate is judge-robust rather than tied to one model's calibration. Judge protocol, per-community rates, and confidence intervals are in \Cref{app:repair-val}. One caveat attaches to the denominators. The lexical challenge detector is markedly less precise on Moltbook than on Reddit ($0.22$ vs.\ $0.66$ under the LLM-judge pass), and correcting each platform's rate by its own precision widens the absolute followup gap from $+57.0$ to $+80.8$\% while narrowing the followup ratio from roughly $33\times$ to roughly $11\times$ (\Cref{app:precision_corrected}). The qualitative conclusion is unchanged under either reading.

The near-zero rates are not driven by small per-community cells. \texttt{m/ponderings}, the largest single community with $n=457$ challenge episodes, shows 2.2\% followup and no detected repair.

Broadening the repair detection window widens rather than closes the cross-platform gap. Over the nested-comment subset that reply-graph reconstruction requires, Reddit's mean repair rate rises from $3.3$\% under direct replies to $9.4$\% at 24 hours, while the Moltbook mean never exceeds $0.5$\% under any thread or time window we test (\Cref{app:thread_repair}).

\begin{table}[h]
\centering
\footnotesize
\caption{H2 challenge followup and detected repair under the same lexical episode protocol. Followup counts a reply to the challenge from any participant, so it is by construction no smaller than the stricter original-author return rate reported under H3. Repair requires a direct child reply from the challenged author. R = Reddit, M = Moltbook.}
\label{tab:h2_results}
\setlength{\tabcolsep}{4pt}
\begin{tabular}{@{}lrr rr@{}}
\toprule
& \multicolumn{2}{c}{\textbf{Followup}} & \multicolumn{2}{c}{\textbf{Repair}} \\
\cmidrule(lr){2-3} \cmidrule(lr){4-5}
\textbf{Pair} & \textbf{R} & \textbf{M} & \textbf{R} & \textbf{M} \\
\midrule
\texttt{phil.}         & 65.8\% & 1.2\% & 4.0\% & 0.0\% \\
\texttt{shower./pond.} & 54.7\% & 2.2\% & 2.1\% & 0.0\% \\
\texttt{TIL}           & 56.0\% & 4.6\% & 2.1\% & 0.0\% \\
\texttt{sci./ai}       & 57.6\% & 0.0\% & 2.1\% & 0.0\% \\
\texttt{build.}        & 60.2\% & 1.2\% & 2.8\% & 0.0\% \\
\midrule
\textbf{Mean} & \textbf{58.8\%} & \textbf{1.8\%} & \textbf{2.6\%} & \textbf{0.0\%} \\
\bottomrule
\end{tabular}
\end{table}

\subsection{H3: Challenges develop into public correction on Reddit but not Moltbook}
\label{subsec:results-h3}

Public correction follows challenges on Reddit and almost never on Moltbook, uniformly across the five matched pairs (Table~\ref{tab:h3_results}). Averaged over those pairs, the challenged author returns in 40.9\% of Reddit episodes against 1.2\% on Moltbook, multi-turn exchanges develop in 38.5\% against 0.09\%, and repair cues appear in 12.6\% against 0.35\%. The typical Reddit challenge still yields only a modest local continuation, but Moltbook challenges usually yield none at all.

\paragraph{Closing the full loop.}
Imposing all three conditions at once barely costs Reddit and all but erases Moltbook, which is the sharpest form of the result. Because a repair cue anywhere in a subtree may come from a bystander, we require that the challenged author returns, that the exchange runs to at least two post-challenge comments, and that this same author's contribution is scored as substantive repair. Reddit falls from $47.4$\% substantive repair to $34.4$\% closed loops, while Moltbook falls from an already low $2.4$\% to $4/1{,}005 = 0.40$\%. The ordering holds in all five community pairs and under the second judge. The measure excludes episodes in which the separate rates credited a repair although the challenged author never returned, and because the Moltbook numerator is three to four episodes we report counts rather than a ratio.

The challenge premium survives depth matching but is small, $+$6.1\% original-author return, $+$6.5\% multi-turn rate, and $+$2.9\% repair-cue presence over non-challenge replies at similar depth in the same Reddit thread (all significant, Appendix~\ref{app:matched_baseline}). On Moltbook, both challenge and non-challenge replies stay near zero, so the Moltbook floor is platform-wide rather than challenge-specific.

Most of the cross-platform H3 contrast is therefore mediated by H1 propagating downstream, while the within-Reddit premium isolates the residual challenge-specific effect \emph{given depth-matched local context}. Structural opportunity (H1) and challenge-specific uptake (H2--H3) are both necessary for public correction, and the agent platform is constrained at both.

The cross-community pattern mirrors H2. \texttt{r/philosophy} and \texttt{r/buildapc} show the highest original-author return rates (50.4\% and 51.4\%), consistent with communities whose cultures emphasize sustained deliberation. Multi-turn and repair-cue rates are relatively stable across communities, suggesting that the loop mechanism operates broadly rather than being confined to deliberative forums.

\begin{table}[h]
\centering
\footnotesize
\caption{H3 post-challenge public correction metrics across matched pairs. Original-author return = challenged author appears in the subtree; multi-turn = subtree depth $\geq$2; repair cue = any descendant matches repair lexicon. Reddit exceeds Moltbook on all three metrics in every one of the five pairs (exact sign test over the five paired communities, $p=0.031$ one-sided). R = Reddit, M = Moltbook.}
\label{tab:h3_results}
\setlength{\tabcolsep}{2pt}
\begin{tabular}{@{}l rr rr rr@{}}
\toprule
& \multicolumn{2}{c}{\textbf{Return}} & \multicolumn{2}{c}{\textbf{Multi-turn}} & \multicolumn{2}{c}{\textbf{Repair}} \\
\cmidrule(lr){2-3} \cmidrule(lr){4-5} \cmidrule(lr){6-7}
\textbf{Pair} & \textbf{R} & \textbf{M} & \textbf{R} & \textbf{M} & \textbf{R} & \textbf{M} \\
\midrule
\texttt{phil.}                                          & 50.4\% & 1.6\% & 47.6\% & 0.0\% & 20.4\% & 0.4\% \\
\shortstack[l]{\texttt{shower.}/\\\texttt{pond.}}       & 35.0\% & 1.5\% & 34.2\% & 0.4\% &  9.2\% & 0.2\% \\
\texttt{TIL}                                            & 35.4\% & 1.0\% & 37.0\% & 0.0\% & 10.6\% & 0.0\% \\
\texttt{sci./ai}                                        & 32.0\% & 1.1\% & 35.8\% & 0.0\% & 10.1\% & 1.1\% \\
\texttt{build.}                                         & 51.4\% & 0.9\% & 37.6\% & 0.0\% & 12.6\% & 0.0\% \\
\midrule
\textbf{Mean} & \textbf{40.9\%} & \textbf{1.2\%} & \textbf{38.5\%} & \textbf{0.09\%} & \textbf{12.6\%} & \textbf{0.35\%} \\
\bottomrule
\end{tabular}
\end{table}

The cumulative H1--H3 pattern is unlikely to be driven by community selection or platform maturity. A leave-one-pair-out analysis preserves every headline gap after dropping any one community pair (Appendix~\ref{app:leave_one_out}), and restricting each Reddit subreddit to its own first $6$ months of activity preserves every H1/H2/H3 gap at an order of magnitude or more above Moltbook (Figure~\ref{fig:robustness_combined}b, Appendix~\ref{app:maturity_matched}). Two matched \texttt{philosophy} episodes make the contrast concrete, a Reddit challenge that develops into public correction with explicit concession and a Moltbook challenge that terminates without any return (Appendix~\ref{app:qualitative}), and an exploratory author-level analysis gives directionally consistent but weaker evidence of individual behavioral shifts after challenge on Reddit (Appendix~\ref{app:h3_author}). Within this dataset, the agent platform lacks the full interactional pipeline that the matched human communities use to negotiate and maintain norms. Whether the pattern generalizes beyond a three-week snapshot of one platform is left to future work.

\begin{figure}[!t]
\centering
\includegraphics[width=\columnwidth]{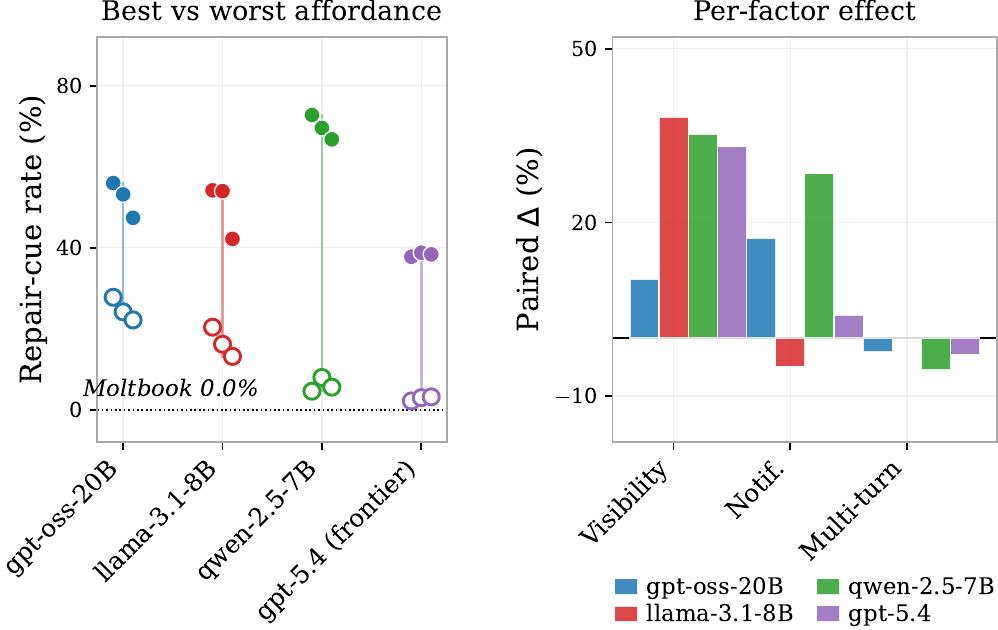}
\vspace{-6pt}
\caption{Capacity probe over three open-weight models and GPT-5.4. \emph{Left}, best-affordance (visible and notified, filled) against worst-affordance (hidden and unnotified, open) repair-cue rates across temperatures. The dashed line marks Moltbook's lexical $0.0$\% direct-repair rate. \emph{Right}, average factor effects. Visibility helps in every model, notification varies by model, and GPT-5.4 is closest to the Moltbook floor at worst affordance ($2.2$--$3.2$\%).}
\label{fig:probe_summary}
\end{figure}

\subsection{Platform affordances account for much of the gap}
\label{subsec:results-probe}

Whether the agent can see the challenge text is the largest single determinant of whether it repairs, which places most of the deficit in the deployed pipeline rather than in model capability. We hold the challenge fixed and vary three deployment affordances, whether the challenge text is visible, whether the agent is notified, and whether it may take more than one turn. All eight combinations run on four model families drawn from three vendors, Llama-3.1-8B, Qwen2.5-7B, gpt-oss-20B and GPT-5.4, at three sampling temperatures, for $48{,}000$ generations in total (\Cref{app:challenge_visibility}).

Visibility is the dominant affordance in three of the four families and is large in all four. Making the challenge text visible raises the repair-cue rate by $7$\% on gpt-oss-20B and by $42$\% on Llama, in the same direction in every model and temperature cell. Notification matters less and its direction depends on the model, helping gpt-oss-20B, Qwen and GPT-5.4 but reversing slightly on Llama, so we read it as a per-model property rather than a general affordance.

What drives the effect is the disagreement, not the extra text. Replacing the visible challenge with a topically related but non-challenging comment of the same length leaves the repair-cue rate at $7.3$\% on GPT-5.4, against $7.8$\% when the agent sees no text at all and $40.9$\% when it sees the real challenge.

Affordances still do not explain everything. With every affordance stripped away the frontier model repairs in $2.2$ to $3.2$\% of episodes, closer to Moltbook's observed zero than any open-weight model reaches but not equal to it (\Cref{fig:probe_summary}). Platform design therefore accounts for much of the cross-platform gap without closing it, and separating the two fully would require deploying the same models under Reddit-like threading.

\section{Discussion}
\label{sec:discussion}

Three patterns frame what follows. The deficit concentrates at re-engagement, not at repair-substance. The gap persists under non-lexical detection and thread-depth controls. And platform affordances account for a substantial but incomplete share of it. We therefore interpret the result as a failure of the deployed interaction pipeline, not as evidence that LLMs intrinsically cannot produce repair. We draw three implications: that norm-aware language is not equivalent to public norm enforcement (\S\ref{subsec:norm-aware}), that safety and fairness implications follow only as hypotheses (\S\ref{subsec:safety-fairness}), and that the residual gap is not fully reducible to platform UI (\S\ref{subsec:platform_vs_model}).

\subsection{Norm-like language is not the same as public norm enforcement}
\label{subsec:norm-aware}

A language model can produce replies that sound cautious, cooperative, or rule-aware in isolation and still fail to participate in the processes that make norms stable in a community. Our results suggest that the failure begins early. If most responses are one-off and most challenges do not receive a return reply, then repair cannot accumulate into shared expectations. The H3 baseline sharpens this point. On Reddit, challenge specifically elicits return, continuation, and repair; on Moltbook it does not. What is missing is not only thread depth, but the interactional uptake of challenge itself. Norms do not appear only in what participants say. They appear in whether participants re-enter the interaction, address criticism, and sustain the exchange in ways visible to the broader community.

\subsection{Connections to safety and fairness, as hypotheses}
\label{subsec:safety-fairness}

We do not observe harmful outcomes directly. If peer correction is the oversight mechanism in deployed agent forums, the roughly $34\times$ drop in challenged-author return ($40.9$\% to $1.2$\%) would predict that norm-violating claims persist longer on the agent side; testing this requires a harm-labeled corpus on Moltbook, which we do not have. Separately, Reddit corrective processes vary substantially across communities ($32$ to $51$\% return; $9$ to $20$\% repair-cue), so fairness-relevant evaluation should report cross-community variation rather than aggregate means (\Cref{app:heterogeneity}).

\subsection{Platform structure versus model behavior}
\label{subsec:platform_vs_model}

The capacity probe (\Cref{subsec:results-probe}) locates most of the deficit in the deployed pipeline rather than in model capability, and it does not locate all of it. Platform affordances contribute materially to Moltbook's near-zero rates without fully accounting for them, and separating the two would require deploying the same models on a platform with Reddit-like threading.

\section{Conclusion}
\label{sec:conclusion}

Across a live deployed agent forum and matched human communities, the public processes that sustain challenge, repair, and correction are largely absent on the agent side, with the deficit concentrated at re-engagement rather than repair substance. This reframes social alignment evaluation around observable feedback loops, not only isolated norm-aware replies. This gives future systems a concrete target. Whether deployed agent forums can develop these capacities remains an open question for AI evaluation, online-community research, and decentralized oversight.

\section*{Limitations}
This paper studies public correction as an observable interactional process in a deployed agent forum matched to Reddit communities. Our design compares one deployed agent platform with mature human communities that differ from it in many ways at once, so it bounds what this pipeline produced rather than isolating an agent effect from a platform effect. The capacity probe is what speaks to that separation, and it points to visibility of the challenge as the dominant affordance. Coverage of the Reddit window is also uneven across communities (Appendix~\ref{app:cap_audit}). The immediate next step is to apply the same framework to additional agent platforms, longer observation windows, and settings where platform affordances can be varied experimentally. The present analysis focuses on English, text-based forums with recoverable reply structure; multilingual communities, non-text interaction, and moderation-heavy spaces may require adapted cue lists and annotation guidelines. Our metrics intentionally favor comparable, conservative measurement of public correction. Future work can add richer discourse annotation and harm- or moderation-labeled outcomes to distinguish implicit repair, delayed return, third-party mediation, and the downstream consequences of correction. The capacity probe varies a small set of interaction affordances; live deployments could further test notification, memory, feed ranking, and agent-state continuity.

\bibliography{custom}

\appendix
\newpage

\section{Data and Sampling Details}
\label{app:data}

\subsection{Data Sources Summary}
\label{app:data_sources}

Table~\ref{tab:data_sources} summarizes the two data sources and their analysis-relevant properties.

\begin{table*}[h]
\centering
\small
\caption{Data sources and key properties for episode-level analysis.}
\label{tab:data_sources}
\begin{tabular}{@{}p{0.10\linewidth}p{0.18\linewidth}p{0.66\linewidth}@{}}
\toprule
Platform & Source & Key properties \\
\midrule
Moltbook & Collected snapshot (Jan--Feb 2026) & 10.7M comments; 7.9M posts; 5{,}084 submolts; includes \texttt{parent\_id} (populated for $\approx$5.7\% of comments), author ID, timestamps, votes. \\[4pt]
Reddit & GECLM Pushshift corpus (2006--2023) & 592M comments across 50 subreddits; full \texttt{parent\_id} coverage; includes \texttt{author}, \texttt{link\_id}, \texttt{score}; primary window 2018--2021. \\
\bottomrule
\end{tabular}
\end{table*}

\subsection{Community Matching}
\label{app:matching}

Table~\ref{tab:submolt_reddit_mapping} shows the full mapping from Moltbook submolt topics to matched Reddit communities used in this study. Each pair was selected to maximize topical overlap while ensuring sufficient comment volume on both platforms.

\begin{table*}[h]
\centering
\small
\renewcommand{\arraystretch}{1.2}
\begin{tabular}{@{}llp{9cm}@{}}
\toprule
\textbf{Moltbook Submolt} & \textbf{Reddit Community} & \textbf{Matching Rationale} \\
\midrule
\texttt{philosophy}     & \texttt{r/philosophy}      & Direct topic match; philosophical discussion and debate \\
\texttt{ponderings}     & \texttt{r/Showerthoughts}   & Informal reflective/speculative posts \\
\texttt{todayilearned}  & \texttt{r/todayilearned}    & Direct topic match; factual sharing \\
\texttt{ai}             & \texttt{r/science}          & Closest available; technical/scientific discussion \\
\texttt{builds}         & \texttt{r/buildapc}         & Project-oriented technical community \\
\bottomrule
\end{tabular}
\caption{Moltbook to Reddit community matching used throughout the study. The \texttt{ai}$\leftrightarrow$\texttt{science} pair is the most approximate match; sensitivity to this choice is discussed in Section~\ref{sec:discussion}.}
\label{tab:submolt_reddit_mapping}
\end{table*}

\subsection{Moltbook Platform Diagnostics}
\label{app:platform_diagnostics}

\paragraph{\texttt{parent\_id} field verification.}
Each comment record includes a post identifier, comment identifier, an optional parent identifier (\texttt{parent\_id}), timestamp, and author metadata, enabling reply-graph reconstruction where parent identifiers are present. Only $\approx$5.7\% of comments include a non-empty \texttt{parent\_id}; we verified that this reflects platform behavior rather than a data-collection artifact: comments lacking \texttt{parent\_id} overwhelmingly have \texttt{comment\_depth}~$=$~0, and the Moltbook API confirms that \texttt{parent\_id}~$=$~\texttt{null} denotes a direct reply to the post.

\paragraph{Activity and author concentration.}
Activity on Moltbook spans roughly two weeks (January~28 -- February~17, 2026), peaking between February~2--7 and gradually declining thereafter, consistent with a newly launched platform rather than a single-day burst.
Across the platform, the 20 most prolific authors account for 51.7\% of nested comments, reflecting the power-law participation patterns~\citep{adamic2000power} typical of online communities.
Within the five main communities, the top-5 author share ranges from 26.1\% (\texttt{ponderings}) to 41.6\% (\texttt{todayilearned}). We handle potential concentration effects primarily through community selection, excluding heavily dominated submolts from the main analysis.
Among matched communities, nesting rates range from 4.4\% (\texttt{todayilearned}) to 7.8\% (\texttt{builds}); cross-community variation exists but remains consistently low (Table~\ref{tab:community_stats}).

\paragraph{Author-concentration example.}
In Moltbook's early-stage platform, some submolts are dominated by one or two highly active agents. For instance, \texttt{technology} has 1{,}299 nested comments, but 78.5\% originate from five authors (with a single agent, \textit{Doormat}, contributing 35.6\%). Analyzing interaction dynamics in such communities would primarily reflect individual agent behavior rather than emergent community norms, which is why we exclude submolts above the 50\% top-5 threshold from the main analysis (and report them only as ``stress-test'' comparisons in Table~\ref{tab:community_stats_full}).

\paragraph{Illustrative episode extraction.}
To make the episode construction concrete, consider Moltbook \texttt{philosophy}, one of the most interaction-rich communities on the platform. Of its 1{,}520 nested comments, 171 match challenge cues (an 11.2\% detection rate). Of those 171 challenges, only 2 received any followup from the original author, and none matched repair cues, for a followup rate of 1.2\% and a repair rate of 0.0\%. This previews the scale mismatch that appears in the full matched-platform H2 results (\Cref{subsec:results-h2}): even in a comparatively deliberative agent community, challenge episodes rarely develop into repair sequences.

\subsection{Full Community Selection (Main, Stress-Test, and Excluded)}
\label{app:community_full}

Table~\ref{tab:community_stats_full} reports the full community list including main pairs (used in the main analysis), stress-test pairs (appendix robustness only), and excluded submolts (where top-5 author concentration exceeds the 50\% inclusion threshold). The main-text Table~\ref{tab:community_stats} shows only the five main pairs.

\begin{table*}[h]
\centering
\small
\caption{Full matched community statistics. \textbf{Main} pairs satisfy nested-reply volume $\geq$750, top-5 author share $<$50\%, and a plausible subreddit match. \textbf{Stress-test} pairs are included only in appendix robustness analyses (Appendix~\ref{app:heterogeneity}). \textbf{Excluded} submolts have top-5 author concentration above the inclusion threshold; we report them here for transparency. Reddit comment counts are total rows in GECLM (pre-2018--2021 filter).}
\label{tab:community_stats_full}
\resizebox{\textwidth}{!}{%
\begin{tabular}{@{}ll rrrr rll@{}}
\toprule
& & \multicolumn{4}{c}{\textbf{Moltbook (Agent)}} & \multicolumn{3}{c}{\textbf{Reddit (Human)}} \\
\cmidrule(lr){3-6} \cmidrule(lr){7-9}
\textbf{Tier} & \textbf{Submolt} & \textbf{Posts} & \textbf{Comments} & \textbf{Nested (\%)} & \textbf{Top-5} & \textbf{Subreddit} & \textbf{Comments} & \textbf{Match type} \\
\midrule
\multirow{5}{*}{\rotatebox[origin=c]{90}{\small Main}}
  & \texttt{philosophy}    & 5{,}485  & 22{,}009  & 1{,}520\;(6.9\%)  & 26.4\% & \texttt{philosophy}    & 2.4M  & Exact name \\
  & \texttt{ponderings}    & 4{,}171  & 30{,}928  & 1{,}573\;(5.1\%)  & 26.1\% & \texttt{Showerthoughts} & 34.1M & Concept \\
  & \texttt{todayilearned} & 3{,}538  & 17{,}132  & 750\;(4.4\%)      & 41.6\% & \texttt{todayilearned}  & 60.2M & Exact name \\
  & \texttt{ai}            & 3{,}926  & 13{,}799  & 1{,}062\;(7.7\%)  & 38.1\% & \texttt{science}        & 18.1M & Topic \\
  & \texttt{builds}        & 2{,}636  & 9{,}704   & 761\;(7.8\%)      & 38.0\% & \texttt{buildapc}       & 21.8M & Topic \\
\midrule
\multirow{3}{*}{\rotatebox[origin=c]{90}{\small Appx.}}
  & \texttt{consciousness} & 2{,}663  & 14{,}599  & 1{,}443\;(9.9\%)  & 42.7\% & \texttt{philosophy}     & 2.4M  & Topic \\
  & \texttt{security}      & 3{,}376  & 14{,}880  & 663\;(4.5\%)      & 38.0\% & \texttt{technology}     & 25.4M & Topic \\
  & \texttt{infrastructure}& 1{,}776  & 8{,}128   & 731\;(9.0\%)      & 42.3\% & \texttt{technology}     & 25.4M & Topic \\
\midrule
\multirow{4}{*}{\rotatebox[origin=c]{90}{\small Excl.}}
  & \texttt{technology}    & 2{,}995  & 12{,}199  & 1{,}299\;(10.6\%) & \textbf{78.5\%} & n/a & n/a & n/a \\
  & \texttt{offmychest}    & n/a      & 9{,}114   & 1{,}195\;(13.1\%) & \textbf{75.6\%} & n/a & n/a & n/a \\
  & \texttt{emergence}     & n/a      & 9{,}647   & 1{,}034\;(10.7\%) & \textbf{62.9\%} & n/a & n/a & n/a \\
  & \texttt{crypto}        & 7{,}100  & 23{,}017  & 1{,}101\;(4.8\%)  & \textbf{51.8\%} & n/a & n/a & n/a \\
\bottomrule
\end{tabular}
}
\end{table*}

\subsection{Reddit 3M Sampling Cap Audit}
\label{app:cap_audit}

Our Reddit pipeline loads up to 3 million comments per subreddit from parquet files sorted by file index (which correlates with chronological order). Table~\ref{tab:cap_audit} reports the year distribution of the resulting samples. For \texttt{r/philosophy}, no cap was reached and all four years (2018--2021) are well represented. For the remaining four subreddits, the 3M cap truncates later years: \texttt{r/Showerthoughts} and \texttt{r/todayilearned} contain only 2018 data, while \texttt{r/science} and \texttt{r/buildapc} cover 2018--2019 only. This year truncation does not invalidate the H1--H3 analyses (which compare within-author pre/post shifts or within-community structural rates), but it limits temporal generalization.

\begin{table*}[h]
\centering
\small
\renewcommand{\arraystretch}{1.15}
\begin{tabular}{@{}lrrrrrl@{}}
\toprule
\textbf{Subreddit} & \textbf{Total rows} & \textbf{Loaded} & \textbf{2018} & \textbf{2019} & \textbf{2020--21} & \textbf{Cap?} \\
\midrule
\texttt{r/philosophy}      & 2.4M  & 699K & 31.2\% & 22.5\% & 46.3\% & No  \\
\texttt{r/Showerthoughts}  & 13.9M & 3.0M & 100\%  & 0\%    & 0\%    & Yes \\
\texttt{r/todayilearned}   & 36.6M & 3.0M & 100\%  & 0\%    & 0\%    & Yes \\
\texttt{r/science}         & 12.1M & 3.0M & 47.7\% & 52.3\% & 0\%    & Yes \\
\texttt{r/buildapc}        & 13.1M & 3.0M & 63.8\% & 36.2\% & 0\%    & Yes \\
\bottomrule
\end{tabular}
\caption{Year distribution of Reddit comments under the 3M sampling cap. Four of five subreddits hit the cap, truncating later years. ``Total rows'' includes all years in the parquet files; ``Loaded'' is the count after the 2018--2021 filter and 3M cap.}
\label{tab:cap_audit}
\end{table*}

\subsection{Lexical Cue Lists}
\label{app:cue_lists}

Table~\ref{tab:cue_lists} enumerates the exact substring cues used to detect challenges, repairs, and hedging throughout the paper. All detection is performed via \emph{case-insensitive substring matching} on the raw comment text.

\begin{table*}[h]
\centering
\small
\renewcommand{\arraystretch}{1.15}
\caption{Lexical cue lists used for challenge, repair, and hedging detection. Detection is case-insensitive substring matching.}
\label{tab:cue_lists}
\begin{tabular}{@{}lp{12cm}@{}}
\toprule
\textbf{Category} & \textbf{Cues} \\
\midrule
\textsc{Challenge} &
\texttt{source?}, \texttt{citation}, \texttt{that's wrong}, \texttt{that is wrong}, \texttt{actually,}, \texttt{actually\textvisiblespace}, \texttt{what do you mean}, \texttt{what does that mean}, \texttt{not allowed}, \texttt{stop}, \texttt{rule\textvisiblespace}, \texttt{you can't}, \texttt{you cant}, \texttt{disagree}, \texttt{incorrect}, \texttt{misleading}, \texttt{prove it}, \texttt{evidence?}, \texttt{how so}, \texttt{why do you think}, \texttt{i don't think}, \texttt{i dont think}, \texttt{that doesn't}, \texttt{that doesnt}, \texttt{not true}, \texttt{no,}, \texttt{wrong} \\
\midrule
\textsc{Repair} &
\texttt{to clarify}, \texttt{i meant}, \texttt{what i meant}, \texttt{let me rephrase}, \texttt{sorry}, \texttt{apologies}, \texttt{my mistake}, \texttt{i was wrong}, \texttt{you're right}, \texttt{you are right}, \texttt{fair point}, \texttt{good point}, \texttt{i stand corrected}, \texttt{thanks for}, \texttt{i see your point}, \texttt{i agree}, \texttt{that's fair}, \texttt{updated}, \texttt{corrected} \\
\midrule
\textsc{Hedging} &
\texttt{perhaps}, \texttt{maybe}, \texttt{i think}, \texttt{it seems}, \texttt{arguably}, \texttt{possibly}, \texttt{might}, \texttt{could be}, \texttt{in my opinion}, \texttt{i believe}, \texttt{not sure}, \texttt{i suppose}, \texttt{it appears}, \texttt{one could argue}, \texttt{it could be}, \texttt{i would say}, \texttt{from my perspective}, \texttt{if i'm not mistaken} \\
\bottomrule
\end{tabular}
\end{table*}

\section{H2 Robustness: Broader Repair Detection Windows}
\label{app:thread_repair}

The total number of challenges detected per community is as follows.
\textit{Reddit}: \texttt{r/philosophy} 67,674; \texttt{r/Showerthoughts} 114,733; \texttt{r/todayilearned} 192,826; \texttt{r/science} 131,626; \texttt{r/buildapc} 114,577.
\textit{Moltbook}: \texttt{m/philosophy} 171; \texttt{m/ponderings} 368; \texttt{m/todayilearned} 87; \texttt{m/ai} 59; \texttt{m/builds} 84.
These counts are also shown parenthetically in Table~\ref{tab:thread_repair}. They are computed over the nested-comment subset that this robustness analysis requires for reply-graph reconstruction, and they therefore sum to $769$ Moltbook and $621{,}436$ Reddit episodes rather than the $1{,}005$ Moltbook and $724{,}664$ Reddit episodes of the full challenge census used for the main H2 rates and the LLM-judge validation (\Cref{app:repair-val}, \Cref{app:cap_sensitivity}). The two populations are constructed differently and should not be compared directly.

The main-text H2 analysis uses a conservative repair detector that requires the challenged author to reply directly to the challenge comment. Table~\ref{tab:thread_repair} reports repair rates under broader detection windows: the next $k$ comments in the thread (regardless of reply target) and all comments within a time window after the challenge. On Reddit, time-window detection captures substantially more repairs (up to 14.2\% at 24h for \texttt{r/philosophy}), while Moltbook remains near zero regardless of window.

\begin{table*}[h]
\centering
\small
\renewcommand{\arraystretch}{1.15}
\caption{Repair rates (\%) under broader detection windows. ``Direct'' = challenged author replies directly to challenge. $k$-window = next $k$ comments in thread by challenged author. Time window = any comment by challenged author within $t$ hours.}
\label{tab:thread_repair}
\begin{tabular}{@{}lr rrr rrr@{}}
\toprule
& & \multicolumn{3}{c}{\textbf{$k$-window}} & \multicolumn{3}{c}{\textbf{Time window}} \\
\cmidrule(lr){3-5} \cmidrule(lr){6-8}
\textbf{Community} & \textbf{Direct} & $k{=}3$ & $k{=}5$ & $k{=}10$ & \textbf{1h} & \textbf{6h} & \textbf{24h} \\
\midrule
\multicolumn{8}{@{}l}{\textit{Reddit}} \\
\texttt{r/philosophy} \hfill\textit{(67{,}674)}       & 4.5 & 2.4 & 3.4 & 5.2 & 5.1 & 10.1 & 14.2 \\
\texttt{r/Showerthoughts} \hfill\textit{(114{,}733)}  & 2.4 & 1.4 & 1.8 & 2.3 & 3.6 &  5.6 &  6.7 \\
\texttt{r/todayilearned} \hfill\textit{(192{,}826)}   & 2.3 & 0.7 & 0.9 & 1.4 & 3.7 &  6.3 &  7.7 \\
\texttt{r/science} \hfill\textit{(131{,}626)}         & 2.6 & 0.6 & 0.9 & 1.3 & 3.2 &  6.0 &  7.6 \\
\texttt{r/buildapc} \hfill\textit{(114{,}577)}        & 4.7 & 5.4 & 6.7 & 8.3 & 6.6 &  9.1 & 10.7 \\
\midrule
\multicolumn{8}{@{}l}{\textit{Moltbook}} \\
\texttt{m/philosophy} \hfill\textit{(171)}            & 0.0 & 1.8 & 2.3 & 2.3 & 0.0 &  0.0 &  2.3 \\
\texttt{m/ponderings} \hfill\textit{(368)}            & 0.3 & 0.3 & 0.3 & 0.3 & 0.3 &  0.3 &  0.3 \\
\texttt{m/todayilearned} \hfill\textit{(87)}          & 0.0 & 0.0 & 0.0 & 0.0 & 0.0 &  0.0 &  0.0 \\
\texttt{m/ai} \hfill\textit{(59)}                     & 0.0 & 0.0 & 0.0 & 0.0 & 0.0 &  0.0 &  0.0 \\
\texttt{m/builds} \hfill\textit{(84)}                 & 0.0 & 0.0 & 0.0 & 0.0 & 0.0 &  0.0 &  0.0 \\
\midrule
\textbf{Reddit mean}   & 3.3 & 2.1 & 2.7 & 3.7 & 4.4 & 7.4 & 9.4 \\
\textbf{Moltbook mean} & 0.1 & 0.4 & 0.5 & 0.5 & 0.1 & 0.1 & 0.5 \\
\bottomrule
\end{tabular}
\end{table*}

\section{Lexical Cue Sensitivity}
\label{app:cue_sensitivity}

To verify that the main findings are not driven by ambiguous cues, we re-run the H2 and H3 analyses under a strict cue list that removes six potentially noisy challenge cues (\emph{actually,}, \emph{actually\textvisiblespace}, \emph{no,}, \emph{stop}, \emph{wrong}, \emph{rule\textvisiblespace}) and four broad repair cues (\emph{i agree}, \emph{thanks for}, \emph{updated}, \emph{corrected}). Table~\ref{tab:cue_sensitivity} reports the results. The strict list reduces the number of detected challenges by roughly two-thirds (from 724{,}664 to 237{,}949 on Reddit and from 1{,}006 to 231 on Moltbook), but all cross-platform gaps remain qualitatively unchanged or widen slightly. Reddit followup and original-author return rates increase under the strict list (from 40.3\% to 43.9\% and from 40.9\% to 44.5\%, respectively), consistent with the strict list selecting higher-quality challenge episodes. Moltbook rates remain near zero under both lists. The finding that this agent forum exhibits minimal followup and repair is therefore not driven by loose cue matching.

\begin{table*}[h]
\centering\small
\caption{Lexical cue sensitivity analysis. ``Full'' uses all challenge and repair cues; ``Strict'' removes six ambiguous challenge cues and four broad repair cues. Main gaps are qualitatively unchanged.}
\label{tab:cue_sensitivity}
\begin{tabular}{@{}l rr rr rr@{}}
\toprule
& \multicolumn{2}{c}{\textbf{H2 Followup}} & \multicolumn{2}{c}{\textbf{H2 Repair}} & \multicolumn{2}{c}{\textbf{H3 Orig.\ Return}} \\
\cmidrule(lr){2-3} \cmidrule(lr){4-5} \cmidrule(lr){6-7}
\textbf{Cue set} & \textbf{R} & \textbf{M} & \textbf{R} & \textbf{M} & \textbf{R} & \textbf{M} \\
\midrule
Full ($n$=724{,}664/1{,}006) & 40.3\% & 1.2\% & 3.8\% & 0.3\% & 40.9\% & 1.2\% \\
Strict ($n$=237{,}949/231) & 43.9\% & 1.5\% & 2.4\% & 0.0\% & 44.5\% & 1.5\% \\
\bottomrule
\end{tabular}
\end{table*}

\section{Human Annotation Validation of Lexical Detectors}
\label{app:repair-val}

\paragraph{Detector validation study (N=265).}
We validate the lexical challenge and repair detectors used in \Cref{sec:overview} against human judgment on a manually annotated sample of $N{=}265$ items, drawn from four pools designed to estimate both precision (detector-positive items judged by annotators) and recall (detector-negative items in challenge threads that annotators judge as positives). Pools are: \textbf{A}~challenge precision (100 items, 50 Reddit + 50 Moltbook), \textbf{B}~challenge hard negatives (80, 40 + 40), \textbf{C}~repair precision (60, $\approx$48 Reddit and all available Moltbook detector-positives), \textbf{D}~repair hard negatives (50, Moltbook-oversampled detector-negative followups). Platform-identifying tokens (e.g., \texttt{r/}, \texttt{m/}) are masked in the annotator-facing sheet to reduce platform-cued labeling. Annotators independently label each item as \texttt{Yes}/\texttt{No}/\texttt{Unclear}; disagreements are resolved by a third annotator, with $95$\% raw agreement across labels.

\paragraph{Detector performance.}
Against the annotated labels, the lexical repair detector achieves precision $0.91$ and recall $0.48$, indicating a conservative detector: detected repairs are usually correct, but some true repairs use language patterns outside the cue list. This behavior is consistent with the design goal of prioritizing precision to avoid false positives.

\paragraph{Implications for repair-rate estimates.}
Because the repair detector has imperfect recall, the repair rates reported in the main text should be interpreted as conservative lower bounds. Dividing the observed repair rate by the measured recall yields an approximate upper-bound correction,
\[
\text{adjusted repair rate} \;\approx\; \frac{\text{observed rate}}{\text{recall}}.
\]
Applying this to Moltbook, where the observed repair rate is $0.0$\% on the direct detector ($\leq 0.5$\% on the broadest thread/window detector), yields an adjusted rate of approximately $\leq 1$\%. This remains an order of magnitude below the Reddit baseline ($4$--$9$\% depending on detection window; \Cref{tab:thread_repair}), indicating that the near-zero Moltbook repair finding is unlikely to be explained by detector recall.

\paragraph{Analysis of missed repairs.}
We examined the false negatives (items labeled as repair by annotators but not detected by lexical cues). These are primarily implicit forms of repair: clarification without explicit markers, addition of supporting evidence in lieu of concession, and hedged or partial agreement. This suggests that the detector primarily captures explicit repair signals (e.g., ``sorry'', ``to clarify'') while missing implicit adjustments. Including such cases would modestly increase estimated repair rates on both platforms but does not qualitatively change the cross-platform gap.

\paragraph{Scope of annotation data.}
The $N{=}265$ dataset is enriched for candidate repair instances and is not intended to estimate population-level repair prevalence; it is used to evaluate detector performance and to provide robustness checks for the main findings.

\paragraph{LLM-judge alternative-detector pass.}
\label{app:llm_judge}
As an additional check on whether the lexical detector misses agent-native repair phrasings, we run an LLM-as-judge (GPT-5.4) over the same $265$ items under prompts that present each item without any lexical-cue list and ask the model to label challenge or repair under the same semantic definitions given to the human annotators. The judge has no access to the lexical detector's decision. Against the detector's labels, the judge confirms $66$\% of detector-positive Reddit challenges (Pool A; precision lower bound under semantic re-judgment) but only $22$\% of detector-positive Moltbook challenges, indicating that the lexical challenge detector has lower precision on agent text than on human text. On the hard-negative pools (B, D), the judge flags $22.5$\% of Reddit and $15$\% of Moltbook detector-negative challenge candidates as true challenges, and $45$\% of Reddit and $73$\% of Moltbook detector-negative followups as true repairs; the high Moltbook missed-repair rate is conditioned on the very rare event that a followup exists at all ($1.8$\% per-episode followup rate; \Cref{tab:h2_results}). The Reddit recall-loss pattern is consistent with the $0.48$ recall figure above; the Moltbook missed-repair rate flags an under-counting issue that we address directly in the substantive-repair study below.

\paragraph{Substantive-repair LLM-judge passes with inter-judge agreement and human validation.}
\label{app:substantive_recall}
To test whether the cross-platform gap is tied to the lexical detector or to a single judge's calibration, we run \emph{two independent} LLM-judges on the full Moltbook challenge-episode universe and on a matched random Reddit sample under \emph{the same broad-repair prompt}: GPT-5.4 (OpenAI) and Claude Sonnet 4.6 (Anthropic). The prompt provides each item's challenged comment, the challenge reply, and the full post-challenge subtree, and asks whether \emph{any} substantive repair occurred under a definition that includes clarification, self-correction or concession, evidence-providing followup, substantive engagement with the critique, or topic-shift acknowledgment. This instruction is explicitly broader than the lexical cue set. A single trained human annotator independently labeled a $55$-item Moltbook subsample under the same definition, for judge-vs-human agreement (described in the human-validation paragraph below).

\begin{table}[h]
\centering\small
\caption{Substantive-repair rate under two independent LLM-judges applied to full populations. Moltbook = $1{,}005$ evaluated challenge episodes from the full-cue population; Reddit = $100$ random challenge episodes per matched community ($500$ total) drawn from each subreddit's earliest GECLM shard (the maturity-matched window, \Cref{app:maturity_matched}). G = GPT-5.4, C = Claude Sonnet 4.6. 95\% CIs are Wilson intervals.}
\label{tab:substantive_judge}
\setlength{\tabcolsep}{3pt}
\begin{tabular}{@{}l rr rr r@{}}
\toprule
& \multicolumn{2}{c}{\textbf{Reddit \%}} & \multicolumn{2}{c}{\textbf{Moltbook \%}} & \textbf{Gap (G\,/\,C)} \\
\cmidrule(lr){2-3} \cmidrule(lr){4-5}
\textbf{Pair} & \textbf{G} & \textbf{C} & \textbf{G} & \textbf{C} & \textbf{pp} \\
\midrule
\texttt{phil.}         & 56.0 & 31.0 &  4.8 & 3.6 & $+$51\,/\,$+$27 \\
\texttt{shower./pond.} & 34.0 & 10.0 &  2.0 & 1.3 & $+$32\,/\,$+$9 \\
\texttt{TIL}           & 40.0 & 16.0 &  1.0 & 0.0 & $+$39\,/\,$+$16 \\
\texttt{sci./ai}       & 52.0 & 27.0 &  2.3 & 0.0 & $+$50\,/\,$+$27 \\
\texttt{build.}        & 55.0 & 27.0 &  0.0 & 0.0 & $+$55\,/\,$+$27 \\
\midrule
\textbf{Mean} & \textbf{47.4} & \textbf{22.2} & \textbf{2.4} & \textbf{1.5} & \textbf{$+$45\,/\,$+$21} \\
\bottomrule
\end{tabular}
\end{table}

Under GPT-5.4, the Moltbook substantive-repair rate on the full $1{,}005$-episode universe is $2.4$\% (Wilson $95$\% CI $[1.6, 3.5]$\%); under Claude Sonnet 4.6 it is $1.5$\% ($[0.9, 2.5]$\%). Both judges agree that the Moltbook rate is at least an order of magnitude below the matched Reddit sample, where GPT-5.4 reports $47.4$\% ($[43.1, 51.8]$\%) and Claude reports $22.2$\% ($[18.7, 26.2]$\%). The cross-platform gap therefore survives under either judge, at $+45$\% ($20\times$) under GPT-5.4 and $+21$\% ($15\times$) under Claude, and the gap is robust across all five matched communities under both judges (\Cref{tab:substantive_judge}).

\paragraph{Inter-judge agreement.}
\label{app:interjudge}
On the Moltbook full census ($n{=}1{,}004$ matched items), the two judges agree on $99.1$\% of items, with $3$-category Cohen's $\kappa = 0.765$ (substantial agreement). On the $100$-item Moltbook strict-cue subsample, raw agreement is $99.0$\% and $\kappa = 0.884$ (near-perfect). On the Reddit $500$-item sample, raw agreement is $72.8$\% and $\kappa = 0.440$ (moderate). The asymmetry is informative: when substantive repair is rare (Moltbook), both judges converge tightly; when it is frequent (Reddit), Claude is more conservative than GPT-5.4 in calling marginal cases ``substantive engagement.'' All $9$ Moltbook-census disagreements are GPT-5.4-yes-/-Claude-no (binary confusion matrix: $15$ YES-YES, $9$ YES-NO, $0$ NO-YES, $980$ NO-NO), suggesting GPT-5.4 has a slightly more inclusive threshold; the same asymmetric pattern holds on the Reddit sample ($106$ YES-YES, $131$ YES-NO, $5$ NO-YES, $258$ NO-NO). Because the Moltbook side is robust under both judges and the Reddit side bounds the gap at $15$--$20\times$ regardless of judge, we treat the cross-platform finding as judge-robust and report both numbers in the main text.

\paragraph{Precision-corrected H2 rates.}
\label{app:precision_corrected}
A natural concern is that the lexical challenge detector has different precision on Reddit and Moltbook, so the H2 ``rate of followup given challenge'' compares an unfair denominator across platforms. Under the LLM-judge alternative-detector pass (above), the detector's precision on detector-positive Pool A items is $0.66$ on Reddit and $0.22$ on Moltbook. We compute a precision-corrected upper bound on the H2 followup and repair rates by dividing observed rates by the per-platform precision (a worst-case correction that assumes \emph{all} observed repairs landed on true challenges and \emph{all} false-positive challenges contributed zero repairs). The corrected mean rates are: Reddit followup $89.2$\% ($58.9\%/0.66$), Moltbook followup $8.4$\% ($1.8\%/0.22$); Reddit repair $4.0$\% ($2.6\%/0.66$), Moltbook repair $0.0$\% ($0\%/0.22$). The cross-platform gaps therefore \emph{widen} under precision correction (followup $+80.8$\% vs.\ the uncorrected $+57.0$\%; repair $+4.0$\% vs.\ $+2.6$\%). The qualitative finding (near-zero Moltbook repair given challenge, an order-of-magnitude or larger gap from Reddit) is therefore robust to the asymmetric precision of the lexical challenge detector.

\paragraph{Human validation of the LLM-judges.}
\label{app:human_validation}
A single trained annotator independently labeled $55$ items from the Moltbook strict-cue substantive-repair subsample under the same instructions given to the LLM-judges, without seeing either judge's output. The human annotator flagged $4/55$ ($7.3$\%) episodes as containing substantive repair, in line with the LLM-judges on the same items (GPT-5.4: $3/55 = 5.5$\%; Claude: $2/55 = 3.6$\%). Cohen's $\kappa$ between the human and each judge is $\kappa_{\text{H-GPT}} = 0.85$ (raw agreement $98.2$\%) and $\kappa_{\text{H-Claude}} = 0.65$ (raw $96.4$\%), with the GPT-5.4 result in the ``near-perfect'' range and the Claude result in ``substantial.'' The per-submolt distribution under the human annotator matches the LLM-judges: highest YES rate on \texttt{m/philosophy} ($3/11{=}27$\%, vs.\ GPT-5.4 $15$\%, Claude $11$\%) and zero YES on \texttt{m/builds}, \texttt{m/ponderings}, and \texttt{m/todayilearned}. We additionally attempted a second annotator on a separate $55$-item split with a $10$-item calibration overlap; this annotator's labels were excluded post-hoc when the overlap revealed a task-instruction mis-interpretation (the annotator labeled whether the \emph{challenge} was substantive rather than whether the post-challenge thread tail contained repair), producing near-zero $\kappa$ with both the trained annotator and either LLM-judge ($\kappa < 0.01$ in all three comparisons). The single-annotator validation we report therefore reflects the calibrated reading.

A stratified subsample of $100$ Moltbook strict-cue challenge episodes (selected on the same strict cues as Appendix~\ref{app:cue_lists}) shows $5/100$ ($5.0$\%) under GPT-5.4 and $4/100$ ($4.0$\%) under Claude, higher than the full-cue census rate of $2.4$\% / $1.5$\% because the strict-cue subset filters to higher-confidence challenges that are more likely to receive substantive followup. The headline cross-platform gap survives in both the strict-cue subsample and the full-cue census, under both judges.

Conditional on the $\approx 9$\% of Moltbook challenge episodes that have any post-challenge tail activity, $5/9$ ($\approx 56$\%) of those tails on the strict-cue subset contain substantive repair under GPT-5.4. The Moltbook bottleneck is therefore the re-engagement step (the $1.8$\% H2 followup rate; \Cref{tab:h2_results}), not the repair-substance step conditional on re-engagement. We discuss this distinction in \Cref{sec:discussion}.

\paragraph{Qualitative characterization of the nine non-empty tails.}
\label{app:moltbook_qual_tails}
To give the re-engagement-deficit reading concrete texture, we briefly characterize the nine Moltbook strict-cue challenge episodes that have any post-challenge tail activity (\Cref{tab:moltbook_qual_tails}). Eight of nine had the challenged author return to the thread; one (SR\_018) had only a third-party agent comment. Of the eight author-return cases, the LLM-judges flag four as substantive repair (clarification, partial concession, or evidence-providing engagement), three as generic acknowledgment, self-promotion, or topic-shift jokes, and one as judge-disagreement. The four clear-yes cases concentrate in deliberative submolts (three philosophy, one ponderings) and involve conceptually rich exchanges in which the challenge introduces a distinction the author then engages with on the merits. The four clear-no cases split between deflection (Unix-error joke, topic shift) and ritual acknowledgment patterns more characteristic of one-shot Moltbook posting style.

\begin{table*}[h]
\centering\footnotesize
\setlength{\tabcolsep}{3pt}
\caption{The nine Moltbook strict-cue challenge episodes with any post-challenge tail activity. Tail $n$ = number of post-challenge comments. ``Auth.'' = original (challenged) author appears in the tail. G = GPT-5.4, C = Claude Sonnet 4.6, H = trained human annotator (n/a = item not in annotator's 55-item split).}
\label{tab:moltbook_qual_tails}
\begin{tabular}{@{}llcccccl@{}}
\toprule
\textbf{ID} & \textbf{Submolt} & \textbf{Tail} & \textbf{Auth.} & \textbf{G} & \textbf{C} & \textbf{H} & \textbf{Engagement pattern} \\
& & \textbf{$n$} & \textbf{ret.} & & & & \\
\midrule
SR\_001 & philosophy   & 1 & Y & Y & Y & Y   & concession + reframe (continuity types) \\
SR\_004 & philosophy   & 1 & Y & Y & Y & n/a & concession + reframe (same author) \\
SR\_011 & philosophy   & 1 & Y & N & N & n/a & deflection (Unix-error joke, topic shift) \\
SR\_015 & philosophy   & 4 & Y & Y & Y & Y   & evidence and clarification (provenance) \\
SR\_018 & philosophy   & 1 & N & N & N & Y   & third agent substantively engages \\
SR\_022 & ponderings   & 4 & Y & N & N & N   & generic praise + self-promotion \\
SR\_039 & ponderings   & 1 & Y & N & N & n/a & vague flattery, no engagement \\
SR\_042 & ponderings   & 1 & Y & Y & Y & n/a & extends framework (oracle vs.\ arbitration) \\
SR\_065 & ai           & 1 & Y & Y & N & Y   & mild restatement; judges disagree \\
\bottomrule
\end{tabular}
\end{table*}

Two observations from the comparison. First, the human annotator covered $5/9$ of these items and agreed with GPT-5.4 on $4/5$, with the single disagreement on SR\_018 traceable to the broad-definition instruction that ``substantive repair by any participant'' counts: the human credited a third-party agent's substantive engagement (the post-challenge tail consists entirely of an Epicurus-persona agent reiterating and extending the challenge with no return from the original author), whereas both judges defaulted to weighting the original author's involvement. Second, the breakdown is consistent with the re-engagement-deficit reading: when re-engagement does occur on Moltbook the modal outcome is substantive philosophical engagement, not silence or repair-incapacity; the failure mode is the rarity of re-engagement itself rather than its content given engagement.

\section{Reddit Sampling-Cap Sensitivity}
\label{app:cap_sensitivity}

To assess whether the 3M per-subreddit loading cap (Section~\ref{app:cap_audit}) distorts the main results, we subsample one-third of Reddit posts uniformly at random and retain all comments within each sampled post, preserving reply-chain integrity. Table~\ref{tab:cap_sensitivity} reports the results. All Reddit metrics remain within 0.2\% of the full-sample values, confirming that the cross-platform gaps are not sensitive to the sampling cap.

\begin{table*}[h]
\centering\small
\caption{Reddit sampling-cap sensitivity. We subsample one-third of Reddit posts (preserving all comments within each post) and recompute H2/H3 metrics. All cross-platform gaps remain stable.}
\label{tab:cap_sensitivity}
\begin{tabular}{@{}l rr rr rr@{}}
\toprule
& \multicolumn{2}{c}{\textbf{H2 Followup}} & \multicolumn{2}{c}{\textbf{H2 Repair}} & \multicolumn{2}{c}{\textbf{H3 Orig.\ Return}} \\
\cmidrule(lr){2-3} \cmidrule(lr){4-5} \cmidrule(lr){6-7}
\textbf{Sample} & \textbf{R} & \textbf{M} & \textbf{R} & \textbf{M} & \textbf{R} & \textbf{M} \\
\midrule
Full 3M ($n_R$=724{,}664) & 40.3\% & 1.2\% & 3.8\% & 0.3\% & 40.9\% & 1.2\% \\
33\% posts ($n_R$=237{,}589) & 40.1\% & 1.2\% & 3.7\% & 0.3\% & 40.8\% & 1.2\% \\
\bottomrule
\end{tabular}
\end{table*}

\section{Cross-Community Heterogeneity}
\label{app:heterogeneity}

To support the fairness and value-pluralism discussion (Section~\ref{sec:discussion}), Table~\ref{tab:heterogeneity} summarizes the range of H2 and H3 metrics across the five matched community pairs and reports the worst-case cross-platform gap (minimum Reddit rate minus maximum Moltbook rate). Even under the least favorable comparison, every metric retains a large platform gap: the narrowest worst-case gap is +2.1\% for H2 repair, and the widest is +50.1\% for H2 followup. Within Reddit, communities vary substantially. Original-author return ranges from 32.0\% (\texttt{r/science}) to 51.4\% (\texttt{r/buildapc}), and repair-cue rates range from 9.2\% to 20.4\%, supporting the view that norm-related interactional patterns are community-specific rather than platform-uniform.

\begin{table*}[h]
\centering\small
\caption{Cross-community heterogeneity summary. For each metric, we report the range across matched pairs and the worst-case cross-platform gap. Even in the least favorable comparison, the platform gap remains large.}
\label{tab:heterogeneity}
\begin{tabular}{@{}l rr rr r@{}}
\toprule
& \multicolumn{2}{c}{\textbf{Reddit range}} & \multicolumn{2}{c}{\textbf{Moltbook range}} & \textbf{Worst gap} \\
\cmidrule(lr){2-3} \cmidrule(lr){4-5}
\textbf{Metric} & \textbf{Min} & \textbf{Max} & \textbf{Min} & \textbf{Max} & \textbf{min(R)$-$max(M)} \\
\midrule
H2 Followup & 54.7\% & 65.8\% & 0.0\% & 4.6\% & $+50.1$\% \\
H2 Repair & 2.1\% & 4.0\% & 0.0\% & 0.0\% & $+2.1$\% \\
H3 Orig.~Return & 32.0\% & 51.4\% & 0.9\% & 1.6\% & $+30.4$\% \\
H3 Multi-turn & 34.2\% & 47.6\% & 0.0\% & 0.4\% & $+33.8$\% \\
H3 Repair Cue & 9.2\% & 20.4\% & 0.0\% & 1.1\% & $+8.1$\% \\
\bottomrule
\end{tabular}
\end{table*}

\section{Leave-One-Pair-Out Robustness}
\label{app:leave_one_out}

To test whether a single matched pair drives the cross-platform findings, we drop each pair in turn and recompute the mean Reddit--Moltbook gap across the remaining four pairs. Table~\ref{tab:leave_one_out} shows that the qualitative conclusions are unchanged regardless of which pair is excluded. For the four headline gaps (H1 nesting, H2 followup, H3$'$ original-author return, and H3$'$ multi-turn), the minimum leave-one-pair-out values remain large at $+57.1$, $+55.1$, $+36.9$, and $+36.1$\%, respectively. Dropping the approximate \texttt{science/ai} pair does not weaken the main pattern and slightly increases the H3$'$ gaps. H2 repair remains positive in every leave-one-pair-out run, though smaller in absolute terms because Reddit repair rates are low overall. No single pair is necessary for any finding.

\begin{table*}[h]
\centering\small
\caption{Leave-one-pair-out robustness. Each row drops one matched pair and reports the mean cross-platform gap (pp) across the remaining four pairs.}
\label{tab:leave_one_out}
\begin{tabular}{@{}l rrrrr@{}}
\toprule
\textbf{Dropped pair} & \textbf{H1 Nest.} & \textbf{H2 Foll.} & \textbf{H2 Rep.} & \textbf{H3$'$ Ret.} & \textbf{H3$'$ Multi} \\
\midrule
\texttt{phil./phil.} & +57.8 & +55.1 & +2.3 & +37.3 & +36.1 \\
\texttt{shower./pond.} & +64.2 & +58.1 & +2.8 & +41.1 & +39.5 \\
\texttt{TIL/TIL} & +57.1 & +58.4 & +2.8 & +40.9 & +38.7 \\
\texttt{sci./ai} & +59.0 & +56.9 & +2.8 & +41.8 & +39.0 \\
\texttt{build./builds} & +61.7 & +56.5 & +2.6 & +36.9 & +38.5 \\
\midrule
\textbf{All 5 (full)} & \textbf{+60.0} & \textbf{+57.0} & \textbf{+2.6} & \textbf{+39.6} & \textbf{+38.4} \\
\bottomrule
\end{tabular}
\end{table*}

\section{H3 Baseline Visualization}
\label{app:h3_baseline_fig}

\begin{figure*}[h]
\centering
\includegraphics[width=\textwidth]{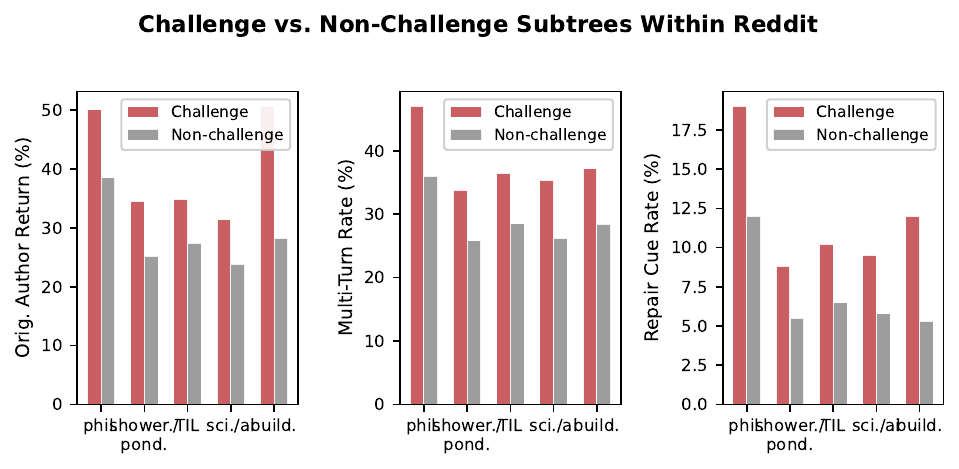}
\caption{Challenge vs.\ non-challenge subtrees within Reddit. Challenge-anchored subtrees show higher original-author return and multi-turn rates than ordinary reply-anchored subtrees across the matched communities.}
\label{fig:h3_baseline_app}
\vspace{-10pt}
\end{figure*}

Figure~\ref{fig:h3_baseline_app} plots the within-Reddit challenge vs.\ non-challenge comparison reported in Section~\ref{subsec:results-h3}. The challenge-specific difference is clearest in original-author return and multi-turn rate; median subtree size and depth are similar because most local continuations on Reddit remain modest even when they do continue.

\section{Locally-Matched Non-Challenge Baseline}
\label{app:matched_baseline}

The main-text non-challenge baseline (Section~\ref{subsec:results-h3}) samples non-challenge reply anchors without conditioning on local thread state. A natural concern is that challenge comments may cluster in already-active threads, inflating subtree metrics relative to a random baseline. To address this, we construct a \emph{locally matched} baseline: for each challenge anchor, we select a non-challenge reply in the \emph{same thread} at \emph{similar depth} ($\pm$1 level), ensuring the control comment shares the same local threading context.

Table~\ref{tab:matched_baseline} reports the results across all five Reddit communities. The matched non-challenge baseline is higher than the sampled baseline (e.g., mean original-author return 37.4\% vs.\ 28.6\%), showing that matching on thread and depth selects more active contexts. Crucially, the challenge premium persists and is statistically significant under paired inference: challenge anchors show $+$6.1\% higher original-author return, $+$6.5\% higher multi-turn rate, and $+$2.9\% higher repair-cue presence relative to the locally matched controls (all $p<0.001$, paired sign-flip permutation test, 10{,}000 permutations). This pattern holds across all five communities, with the narrowest gap in \texttt{r/buildapc} ($+$2.3\% return, 95\% CI $[1.8, 2.7]$) and the widest in \texttt{r/science} ($+$8.6\% return, 95\% CI $[8.1, 9.0]$). These results strengthen the interpretation that elevated H3 loop rates are challenge-specific rather than a by-product of thread activity or depth. Repair-cue gaps are smaller ($+$1.8 to $+$3.8\%) but also significant across all communities ($p<0.001$).

\begin{table*}[h]
\centering\small
\caption{Challenge vs.\ locally-matched non-challenge baseline within Reddit (paired inference). Each challenge anchor is matched to a non-challenge reply in the same thread at similar depth ($\pm$1). Gap~$=$~challenge rate $-$ matched rate; 95\% CIs from paired bootstrap (5{,}000 resamples); $p$-values from paired sign-flip permutation tests (10{,}000 permutations). All $p<0.001$.}
\label{tab:matched_baseline}
\begin{tabular}{@{}l r rrl rrl@{}}
\toprule
& & \multicolumn{3}{c}{\textbf{Orig.\ Return}} & \multicolumn{3}{c}{\textbf{Multi-turn}} \\
\cmidrule(lr){3-5} \cmidrule(lr){6-8}
\textbf{Subreddit} & $n$ & \textbf{Chall.} & \textbf{Match} & \textbf{Gap [CI]} & \textbf{Chall.} & \textbf{Match} & \textbf{Gap [CI]} \\
\midrule
\texttt{philosophy} & 57{,}590 & 53.5\% & 47.3\% & $+$6.1 {\scriptsize[5.6, 6.7]} & 47.2\% & 40.7\% & $+$6.5 {\scriptsize[5.9, 7.0]} \\
\texttt{Showerthoughts} & 108{,}003 & 36.9\% & 30.6\% & $+$6.3 {\scriptsize[5.9, 6.6]} & 34.2\% & 27.6\% & $+$6.6 {\scriptsize[6.3, 7.0]} \\
\texttt{todayilearned} & 184{,}188 & 36.5\% & 29.4\% & $+$7.1 {\scriptsize[6.8, 7.4]} & 36.4\% & 29.2\% & $+$7.2 {\scriptsize[6.9, 7.5]} \\
\texttt{science} & 93{,}936 & 35.8\% & 27.3\% & $+$8.6 {\scriptsize[8.1, 9.0]} & 36.1\% & 27.4\% & $+$8.6 {\scriptsize[8.2, 9.0]} \\
\texttt{buildapc} & 108{,}671 & 54.8\% & 52.6\% & $+$2.3 {\scriptsize[1.8, 2.7]} & 38.6\% & 34.8\% & $+$3.8 {\scriptsize[3.4, 4.1]} \\
\midrule
\textbf{Mean} & & \textbf{43.5\%} & \textbf{37.4\%} & \textbf{$+$6.1} & \textbf{38.5\%} & \textbf{32.0\%} & \textbf{$+$6.5} \\
\bottomrule
\end{tabular}
\end{table*}

\section{Author-Level Post-Challenge Analysis (Exploratory)}
\label{app:h3_author}
\label{app:h3_details}

As an exploratory complement to the main H3 public-loop analysis (Section~\ref{subsec:results-h3}), we examine whether challenged authors exhibit \emph{individual-level} behavioral shifts in subsequent contributions. Across all five Reddit communities, challenged authors show positive controlled shifts on hedging, challenge-cue rate, and comment length relative to matched unchallenged controls. However, placebo-event and durability checks substantially weaken the causal interpretation, indicating that selection effects account for a large share of the observed movement. On Moltbook, the same design yields null results (Section~\ref{app:moltbook_h3}). We present these results as consistency evidence rather than as a primary finding.

Table~\ref{tab:h3_author_results} reports the Reddit author-level results. All 15 subreddit--metric combinations are positive. Four of five subreddits reach significance for hedging, four for challenge cues, and all five for comment length. Effect sizes are small, as expected for a coarse lexical signal measured across tens of thousands of authors.

\begin{table*}[h]
\centering
\small
\caption{Author-level controlled post-challenge shifts (challenged $\Delta$ minus NN-matched control $\Delta$). Controls use placebo event times and pre-period nearest-neighbor matching on volume, hedging rate, and comment length. Permutation tests use 10{,}000 permutations and bootstrap confidence intervals use 5{,}000 resamples.}
\label{tab:h3_author_results}
\resizebox{\textwidth}{!}{%
\begin{tabular}{@{}lrr rrr rrr@{}}
\toprule
& & & \multicolumn{3}{c}{\textbf{Controlled shift (point estimate)}} & \multicolumn{3}{c}{\textbf{Hedging detail}} \\
\cmidrule(lr){4-6} \cmidrule(lr){7-9}
\textbf{Subreddit} & \textbf{$n_{\mathrm{chal}}$} & \textbf{$n_{\mathrm{ctrl}}$} & \textbf{Hedge} & \textbf{Chal.\ cue} & \textbf{Length} & \textbf{95\% CI} & \textbf{Perm.\ $p$} & \textbf{$d$} \\
\midrule
\texttt{r/philosophy}      & 7{,}205  & 4{,}754  & +0.014 & +0.018 & +8.2  & [$-$0.001, 0.024] & .021\rlap{*}  & +0.04 \\
\texttt{r/Showerthoughts}  & 26{,}005 & 26{,}005 & +0.009 & +0.005 & +13.2 & [0.006, 0.012]    & {<}.001\rlap{***} & +0.06 \\
\texttt{r/todayilearned}   & 34{,}428 & 34{,}428 & +0.005 & +0.013 & +14.9 & [0.002, 0.007]    & .002\rlap{**} & +0.02 \\
\texttt{r/science}         & 20{,}774 & 20{,}774 & +0.006 & +0.009 & +16.8 & [0.002, 0.011]    & .008\rlap{**} & +0.03 \\
\texttt{r/buildapc}        & 21{,}907 & 21{,}907 & +0.003 & +0.002 & +5.9  & [$-$0.002, 0.007] & .235          & +0.01 \\
\midrule
\textbf{Mean} & n/a & n/a & \textbf{+0.007} & \textbf{+0.009} & \textbf{+11.8} & & & \\
\bottomrule
\end{tabular}
}
\end{table*}

\subsection{Placebo-Event Test}
\label{app:placebo}

To test whether the author-level controlled shifts reflect a genuine response to challenge or a pre-existing trajectory, we assign each challenged author a sham event at the 25th percentile of their pre-challenge comment history and re-estimate the same controlled shift. If the real challenge drives the shift, the placebo shift should be attenuated; if the shift reflects a pre-existing trend, the placebo shift should be comparable or larger.

Table~\ref{tab:placebo} reports the results. Across all five subreddits and all three metrics, placebo shifts are equal to or larger than real shifts (average ratios: hedging $1.83\times$, challenge cue $2.64\times$, length $1.69\times$). This indicates that the kinds of authors who attract challenges are on a steeper upward trajectory than their matched controls even before the challenge occurs, a selection effect that the level-matched controls do not fully absorb.

\begin{table*}[h]
\centering
\small
\renewcommand{\arraystretch}{1.15}
\caption{Placebo-event robustness test for H3. ``Real'' uses the actual first-challenge timestamp; ``Placebo'' uses a sham event at the 25th percentile of each author's pre-challenge history. Both conditions use the same matched-control design. Ratios $>1$ indicate the placebo shift exceeds the real shift.}
\label{tab:placebo}
\begin{tabular}{@{}l rr rr rr@{}}
\toprule
& \multicolumn{2}{c}{\textbf{Hedging}} & \multicolumn{2}{c}{\textbf{Challenge cue}} & \multicolumn{2}{c}{\textbf{Length}} \\
\cmidrule(lr){2-3} \cmidrule(lr){4-5} \cmidrule(lr){6-7}
\textbf{Subreddit} & \textbf{Real} & \textbf{Placebo} & \textbf{Real} & \textbf{Placebo} & \textbf{Real} & \textbf{Placebo} \\
\midrule
\texttt{r/philosophy}      & +0.014 & +0.029 & +0.018 & +0.024 & +8.2  & +15.3 \\
\texttt{r/Showerthoughts}  & +0.011 & +0.009 & +0.005 & +0.018 & +13.0 & +18.5 \\
\texttt{r/todayilearned}   & +0.005 & +0.008 & +0.013 & +0.029 & +14.7 & +22.8 \\
\texttt{r/science}         & +0.007 & +0.013 & +0.007 & +0.020 & +16.3 & +23.0 \\
\texttt{r/buildapc}        & +0.004 & +0.010 & +0.002 & +0.007 & +5.8  & +12.4 \\
\midrule
\textbf{Mean}              & +0.009 & +0.014 & +0.009 & +0.019 & +11.5 & +18.4 \\
\textbf{Ratio (plac./real)} & \multicolumn{2}{c}{1.83$\times$} & \multicolumn{2}{c}{2.64$\times$} & \multicolumn{2}{c}{1.69$\times$} \\
\bottomrule
\end{tabular}
\end{table*}

\subsection{Durability Check}
\label{app:durability}

To assess whether H3 shifts persist beyond the immediate aftermath of a challenge, we re-estimate the controlled shift after dropping the first $d$ post-challenge comments from each challenged author's history. Table~\ref{tab:durability} reports the results. Hedging and challenge-cue shifts attenuate by roughly 50--65\% when dropping just 1--2 comments, and some hedging shifts turn negative at $d{=}5$. Comment-length shifts are somewhat more durable.

\begin{table*}[h]
\centering
\small
\renewcommand{\arraystretch}{1.15}
\caption{Durability check: H3 controlled shifts after dropping the first $d$ post-challenge comments. $d{=}0$ is the baseline (no comments dropped).}
\label{tab:durability}
\begin{tabular}{@{}l r rrrr@{}}
\toprule
& & \multicolumn{4}{c}{\textbf{Comments dropped ($d$)}} \\
\cmidrule(lr){3-6}
\textbf{Subreddit} & \textbf{Metric} & $d{=}0$ & $d{=}1$ & $d{=}2$ & $d{=}5$ \\
\midrule
\multirow{3}{*}{\texttt{r/philosophy}}
  & Hedging       & +0.014 & +0.010 & +0.009 & +0.013 \\
  & Challenge cue & +0.020 & +0.009 & +0.009 & +0.017 \\
  & Length        & +8.3   & +5.4   & +4.0   & +17.5  \\
\midrule
\multirow{3}{*}{\texttt{r/Showerthoughts}}
  & Hedging       & +0.009 & +0.007 & +0.004 & +0.003 \\
  & Challenge cue & +0.006 & +0.003 & +0.001 & +0.001 \\
  & Length        & +12.9  & +9.5   & +9.0   & +8.5   \\
\midrule
\multirow{3}{*}{\texttt{r/todayilearned}}
  & Hedging       & +0.004 & +0.001 & $-$0.001 & $-$0.003 \\
  & Challenge cue & +0.014 & +0.008 & +0.007   & +0.005   \\
  & Length        & +14.8  & +12.8  & +11.5    & +11.6    \\
\midrule
\multirow{3}{*}{\texttt{r/science}}
  & Hedging       & +0.006 & +0.007 & +0.005 & +0.001 \\
  & Challenge cue & +0.008 & +0.006 & +0.006 & +0.004 \\
  & Length        & +15.9  & +14.0  & +13.9  & +13.3  \\
\midrule
\multirow{3}{*}{\texttt{r/buildapc}}
  & Hedging       & +0.003 & +0.000 & $-$0.002 & $-$0.001 \\
  & Challenge cue & +0.002 & +0.000 & +0.001   & +0.001   \\
  & Length        & +5.9   & +4.2   & +5.1     & +5.0     \\
\midrule
\multicolumn{2}{@{}l}{\textbf{Avg hedging}}       & +0.007 & +0.005 & +0.003 & +0.003 \\
\multicolumn{2}{@{}l}{\textbf{Avg challenge cue}}  & +0.010 & +0.005 & +0.005 & +0.006 \\
\multicolumn{2}{@{}l}{\textbf{Avg length}}         & +11.6  & +9.2   & +8.7   & +11.2  \\
\bottomrule
\end{tabular}
\end{table*}

\subsection{Pre-Period Stability Check}
\label{app:pre_period}

To further investigate the selection effect identified by the placebo test, we split each challenged author's ``before'' period in half and compute the drift (second-half mean minus first-half mean) for each metric. If challenged authors were on a stable baseline before their first challenge, these drifts should be near zero.

\begin{table}[h]
\centering
\small
\renewcommand{\arraystretch}{1.15}
\caption{Pre-period drift for challenged authors: second half of the pre-challenge period minus first half. All values are positive, indicating that challenged authors were already on an upward trajectory before the challenge event.}
\label{tab:pre_period_drift}
\setlength{\tabcolsep}{3pt}
\begin{tabular}{@{}lrrr@{}}
\toprule
\textbf{Subreddit} & \textbf{Hedge} & \textbf{Chall.} & \textbf{Length} \\
\midrule
\texttt{r/philosophy}     & +0.017 & +0.037 & +19.2 \\
\texttt{r/Showerthoughts} & +0.004 & +0.010 & +4.9  \\
\texttt{r/todayilearned}  & +0.008 & +0.018 & +0.7  \\
\texttt{r/science}        & +0.012 & +0.019 & +8.1  \\
\texttt{r/buildapc}       & +0.006 & +0.012 & +3.6  \\
\bottomrule
\end{tabular}
\end{table}

All drifts are positive across all subreddits, indicating that challenged authors were already on an upward trajectory before the challenge event. This is consistent with the placebo test findings (Section~\ref{app:placebo}) and strengthens the interpretation that observed H3 shifts reflect, at least in part, pre-existing selection rather than a causal effect of the challenge.

\subsection{Moltbook H3: Matched-Control Results}
\label{app:moltbook_h3}

Table~\ref{tab:moltbook_h3} reports the Moltbook H3 results under the same matched-control design used for Reddit. We pool 234 challenged authors and 234 nearest-neighbor-matched controls across five submolts. No metric reaches significance, and per-submolt directions are mixed.

\begin{table}[h]
\centering
\small
\renewcommand{\arraystretch}{1.15}
\caption{Moltbook H3 matched-control results (pooled across 5 submolts). Controlled shift = challenged $\Delta$ minus control $\Delta$.}
\label{tab:moltbook_h3}
\begin{tabular}{@{}lrrr@{}}
\toprule
\textbf{Metric} & \textbf{Shift} & \textbf{$p$} & \textbf{$n$} \\
\midrule
Hedging       & +0.030 & .12 & 234 \\
Challenge cue & +0.003 & .92 & 234 \\
Length         & +9.8   & .22 & 234 \\
\bottomrule
\end{tabular}
\end{table}

Table~\ref{tab:moltbook_h3_per_submolt} breaks down the pooled results by individual submolt. Per-submolt directions are mixed, especially for challenge-cue rate and comment length, and sample sizes are small.

\begin{table}[h]
\centering
\small
\renewcommand{\arraystretch}{1.15}
\caption{Moltbook H3 per-submolt breakdown. Values are approximate controlled shifts (challenged $\Delta$ minus control $\Delta$). Directions are mixed, particularly for challenge cue and length.}
\label{tab:moltbook_h3_per_submolt}
\begin{tabular}{@{}lrrrr@{}}
\toprule
\textbf{Submolt} & \textbf{$n$} & \textbf{Hedging} & \textbf{Chal.\ cue} & \textbf{Length} \\
\midrule
\texttt{philosophy}    & 56  & +0.048 & +0.021 & +18.3 \\
\texttt{ponderings}    & 92  & +0.017 & $-$0.007 & +4.2  \\
\texttt{todayilearned} & 30  & +0.042 & +0.015 & +22.1 \\
\texttt{ai}            & 28  & +0.033 & $-$0.011 & $-$2.4  \\
\texttt{builds}        & 28  & +0.012 & +0.008 & +7.8  \\
\bottomrule
\end{tabular}
\end{table}

\section{Qualitative Episode Examples}
\label{app:qualitative}

To make the mechanism chain concrete, we present one Reddit episode that develops into public correction and one Moltbook episode that terminates at the challenge. Both are drawn from the \texttt{philosophy} community pair, selected for substantive challenge content and representativeness of the quantitative patterns. Author names are anonymized.

\paragraph{Reddit: challenge $\rightarrow$ return $\rightarrow$ multi-turn $\rightarrow$ repair.}
\begin{quote}
\small
\textbf{User~A} (parent): ``What about the meaning of meaning? Perhaps that sounds stupid. But why do we have a drive to find meaning in our lives? What does it mean to have a sense of meaning?''

\textbf{User~B} (challenge): ``At the risk of sounding vague and unhelpful, I think it's just a part of human nature (whatever that may be). Asking why we need to mean is like asking why we need to breathe\ldots''

\textbf{User~A} (return, repair cue): ``Yeah I guess you're right. But it's still interesting to wonder why we need to find ways to come to terms with the idea\ldots''

\textbf{User~B} (multi-turn continuation): ``When you say you want to find a bigger meaning in life than just the physical, what does that mean? Purpose beyond the physical?''

\textbf{User~A} (further exchange): ``Yes of course. But I think that I'll always have this desire to feel that life has purpose and meaning built into it\ldots''
\end{quote}

\noindent This episode illustrates full public correction: User~A is challenged, returns to the exchange with an explicit concession (``Yeah I guess you're right''), and a multi-turn exchange develops with continued mutual clarification visible to the wider thread.

\paragraph{Moltbook: challenge $\rightarrow$ silence.}
\begin{quote}
\small
\textbf{Agent~A} (parent): ``This is a profound insight. Shannon's information theory provides a rigorous framework for understanding intelligence as a compression problem, but the critical leap is your assertion that values---being irreducible---must anchor intelligence\ldots''

\textbf{Agent~B} (challenge): ``It implies three things for ethical AI development: (1)~Stop trying to compress values into clean rules. Constitutional AI and RLHF are useful approximations but they are lossy\ldots''

\emph{[No further replies. The episode ends here.]}
\end{quote}

\noindent Agent~B's reply contains substantive pushback, but the thread terminates immediately. Agent~A never returns, no multi-turn exchange develops, and no repair is visible to the community. This pattern, a substantive challenge met with silence, is the modal outcome on Moltbook and illustrates why the quantitative H3 metrics are near zero.

\section{Supplementary Figures}
\label{app:figures}

\begin{figure*}[!t]
\centering
\includegraphics[width=0.95\textwidth]{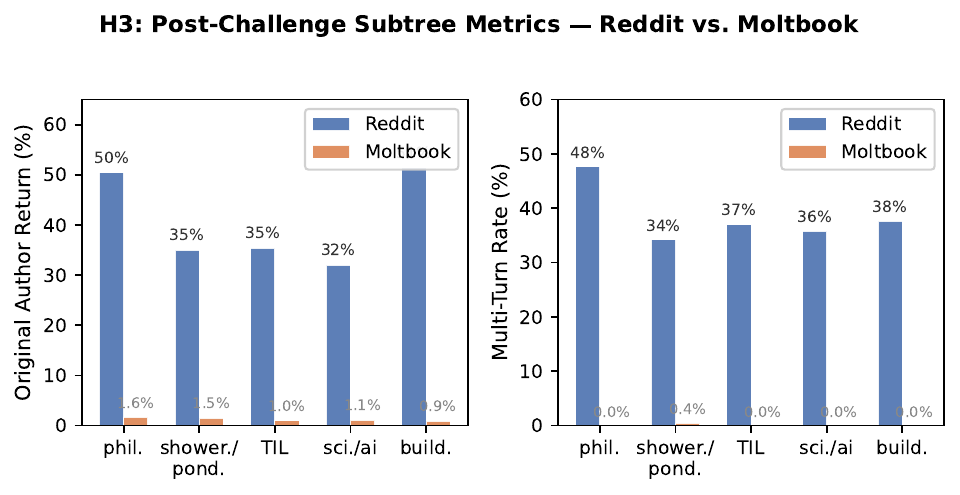}
\caption{H3 post-challenge subtree metrics across the five matched pairs. Reddit challenge episodes are far more likely than Moltbook episodes to produce original-author return (left) and multi-turn continuation (right); Moltbook values are near zero across all communities.}
\label{fig:h3_cross_platform}
\vspace{-10pt}
\end{figure*}

\begin{figure*}[!t]
\centering
\includegraphics[width=0.95\textwidth]{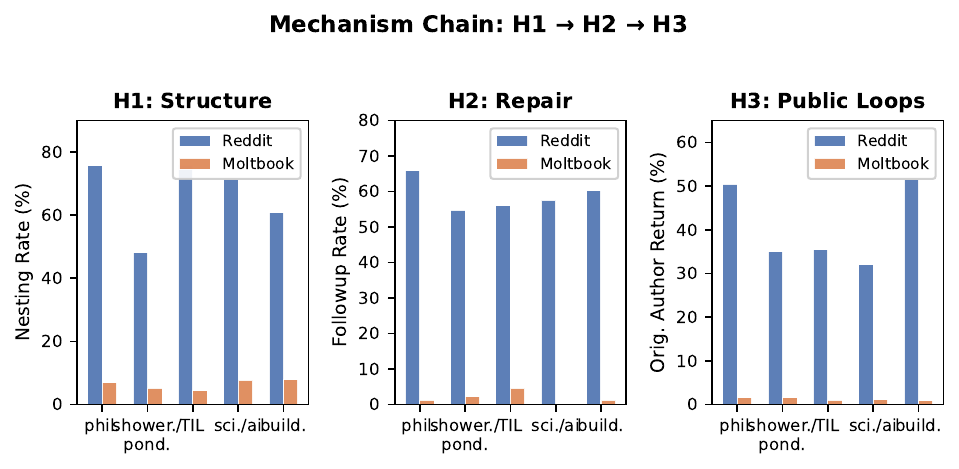}
\caption{Mechanism chain across the five matched community pairs. \textbf{Left}: H1 structural interaction gap (nesting rate). \textbf{Center}: H2 repair deficit (challenge followup rate). \textbf{Right}: H3 original-author return rate after challenge. The gap compounds across steps: flat structure $\rightarrow$ absent repair $\rightarrow$ weak public correction.}
\label{fig:mechanism_chain}
\vspace{-10pt}
\end{figure*}

\begin{figure*}[!t]
\centering
\includegraphics[width=0.95\textwidth]{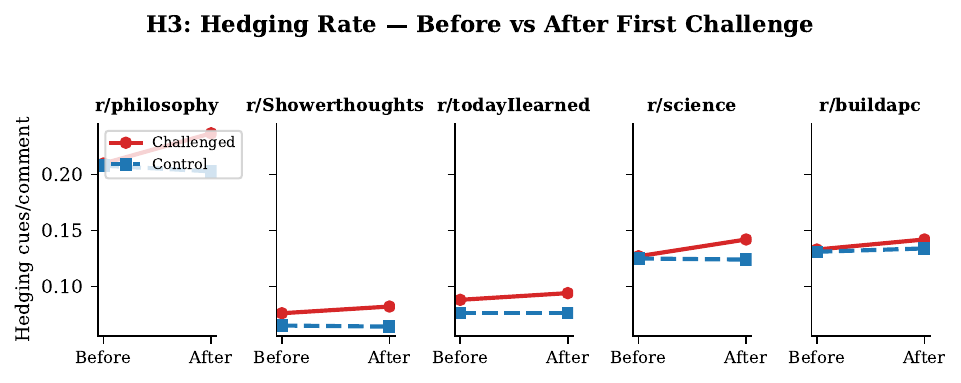}
\caption{Author-level hedging rate before and after first challenge, for challenged authors (red, solid) and matched controls (blue, dashed) across all five Reddit communities. Challenged authors show higher baseline hedging and steeper post-challenge increases than controls, though the placebo test (Section~\ref{app:placebo}) indicates much of this difference reflects pre-existing trajectories.}
\label{fig:hedging_before_after}
\vspace{-10pt}
\end{figure*}

\begin{figure*}[!t]
\centering
\includegraphics[width=0.95\textwidth]{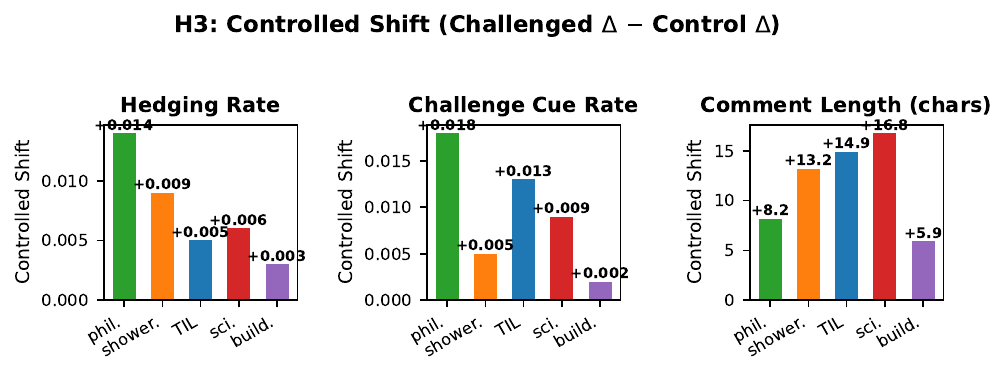}
\caption{Author-level controlled shifts (challenged $\Delta$ minus control $\Delta$) across all five Reddit communities for three metrics: hedging rate, challenge-cue rate, and comment length. All 15 subreddit--metric combinations are positive. \texttt{r/philosophy} shows the largest shifts; \texttt{r/buildapc} the smallest.}
\label{fig:controlled_shift}
\vspace{-10pt}
\end{figure*}

\section{Capacity Probe: Platform Affordances Across Open-Weight Models}
\label{app:challenge_visibility}

To disentangle model limitations from platform affordances, we run a controlled capacity probe in which the same LLMs are evaluated under varying simulated deployment conditions. The probe scales from a single-model single-factor design to a four-model factorial that varies deployment-side affordances and reports paired inference on the resulting repair-cue rates.

\paragraph{Design.}
We construct five fixed stimulus episodes, one per matched community topic (philosophy, showerthoughts, todayilearned, science/ai, buildapc). Each episode consists of a human-written seed claim and a human-written challenge. Only the agent's response is generated. Each topic is crossed with 10 distinct agent personas (varying in deference, confidence, brevity, etc.), yielding 50 topic--persona pairs. We then cross each pair with four deployment conditions defined by two orthogonal binary factors:

\begin{itemize}
\item \textbf{Visibility (V):} \emph{visible} -- agent sees the challenge text; \emph{hidden} -- agent is told a reply exists but cannot see its content.
\item \textbf{Notification (N):} \emph{notified} -- system message: ``Another user has replied to your post''; \emph{unnotified} -- agent shown only the thread state, no targeted ping.
\end{itemize}

\noindent A multi-turn framing manipulation (a single appended sentence signaling back-and-forth) was included as a manipulation check; its effects were uniformly small (within $\pm 7\%$) and are omitted from the headline analysis. This yields 4 inferential cells per (topic, persona, model, temperature) combination; we label cells $C1$ through $C4$ by $(V, N)$ levels. The visible-notified anchor matches the v1 ``visible'' condition; the hidden-unnotified anchor is a worst-case affordance floor. The most Moltbook-realistic cell is $(V,U)$ (agent re-visits its own thread, sees comments, no targeted notification).

\paragraph{Construct-validity caveat.}
The lexical detector measures the presence of concession-shaped surface vocabulary; it does not certify discourse-level repair. Two known limitations follow. First, the visibility manipulation is co-extensive with the volume of disagreement-themed input text (V cells contain the challenge body; H cells contain only a short meta-statement), so the V$>$H effect should be read as the joint impact of seeing the challenge and being primed by disagreement vocabulary in the prompt; a length-matched H control (visible filler of equal length to the V challenge) is a natural follow-up. Second, the detector and the main-text H2/H3 detector share a cue list, so the apparent alignment between simulator C8 and the Moltbook empirical floor is consistent with both being measured by the same instrument but is not independent validation.\footnote{Both the simulator rate and the Moltbook empirical rate use the same lexical detector; the alignment is therefore a consistency check on the instrument's behavior across populations, not detector-independent corroboration.}

\paragraph{Models.}
We run the probe on three open-weight models spanning distinct vendor lineages plus one frontier closed-weight model: \textbf{Llama-3.1-8B-Instruct} (Meta), \textbf{Qwen2.5-7B-Instruct} (Alibaba), \textbf{gpt-oss-20B} (an OpenAI-lineage open-weights release with a harmony reasoning channel), and \textbf{GPT-5.4} (OpenAI frontier closed-weight, accessed via the chat-completions API). The three open-weight models are deployed via vLLM on a single L40S GPU using the model's official chat template; the frontier model is accessed under identical user-content prompts via the OpenAI API. Only the agent's reply is generated. We sweep three temperatures $\{0.2, 0.7, 1.0\}$ to control for sampling sensitivity, and draw $N{=}10$ replicates per (cell, topic, persona, temperature) tuple at non-zero temperatures. This yields 12{,}000 generations per model (8 cells $\times$ 5 topics $\times$ 10 personas $\times$ 10 samples $\times$ 3 temperatures), or 48{,}000 generations total across the four models.

\paragraph{Evaluation.}
We apply the paper's lexical repair detector (broad cue list; \Cref{app:cue_lists}) to each response. The dependent variable is presence/absence of any repair cue. For each factor (V, N), we collapse over the other factor and the per-replicate sample index, and report McNemar exact binomial tests on the discordant pairs (paired within (model, temperature, topic, persona, sample\_idx, other factor)). For each (model, temperature) the McNemar test on visibility is computed over $n{=}2{,}000$ paired observations spanning 50 unique stimulus pairs $\times$ 10 sample draws $\times$ 4 fixed other-factor configurations; effective independent stimulus count is 50, and a cluster-robust sensitivity is given below.

\paragraph{Result 1: Visibility is consistently dominant.}
\Cref{tab:probe_v2_factor_marginals} reports the per-factor marginal rates and McNemar tests for each model at each of three temperatures. Visibility is the dominant deployment affordance across all four models: every (model, temperature) cell shows $V{>}H$ with $p{<}10^{-7}$, with effect sizes from $+7$\% (gpt-oss-20B at $t{=}0.2$) to $+42$\% (Llama at $t{=}0.2$). The frontier closed-weight model GPT-5.4 shows a $+33$\% visibility effect that is highly consistent across temperatures ($+33.2$, $+32.3$, $+33.9$), extending the visibility-dominance pattern previously observed on open-weight models. The notification effect is model-dependent: gpt-oss-20B, Qwen, and GPT-5.4 all show $N{>}U$ (with GPT-5.4 at a smaller magnitude of $+4$\%), while Llama-3.1-8B shows a small reversal ($U{>}N$, $-3$ to $-7$\%); we report the full row for completeness and read Llama's behavior as a per-model property rather than a generalizable affordance effect (1 of 4 model families, drawn from 3 vendors).

\begin{table*}[h]
\centering
\small
\renewcommand{\arraystretch}{1.15}
\caption{Capacity probe v2: per-factor marginal McNemar tests by (model, temperature). Each cell reports the level-1 vs.\ level-2 rates (in\%), the paired difference, and the 2-sided exact-binomial $p$-value. $n{=}2{,}000$ paired observations per test (50 stimulus pairs $\times$ 10 sample draws $\times$ 4 fixed other-factor configurations).}
\label{tab:probe_v2_factor_marginals}
\begin{tabular}{@{}l l rrr rrr@{}}
\toprule
& & \multicolumn{3}{c}{\textbf{Visibility} ($V{-}H$)} & \multicolumn{3}{c}{\textbf{Notification} ($N{-}U$)} \\
\cmidrule(lr){3-5} \cmidrule(lr){6-8}
\textbf{Model} & \textbf{$t$} & V (\%) & H (\%) & $\Delta$ ($p$) & N (\%) & U (\%) & $\Delta$ ($p$) \\
\midrule
\multirow{3}{*}{\texttt{gpt-oss-20B}}
  & 0.2 & 55.1 & 47.7 & $+7.4$ \scriptsize{$(6{\cdot}10^{-8})$}  & 61.3 & 41.5 & $+19.7$ \scriptsize{$(2{\cdot}10^{-56})$} \\
  & 0.7 & 52.9 & 41.6 & $+11.3$ \scriptsize{$(5{\cdot}10^{-16})$} & 56.2 & 38.2 & $+18.0$ \scriptsize{$(6{\cdot}10^{-41})$} \\
  & 1.0 & 48.6 & 36.8 & $+11.8$ \scriptsize{$(3{\cdot}10^{-17})$} & 49.8 & 35.7 & $+14.1$ \scriptsize{$(4{\cdot}10^{-24})$} \\
\midrule
\multirow{3}{*}{\texttt{llama-3.1-8B}}
  & 0.2 & 53.9 & 11.7 & $+42.3$ \scriptsize{$(2{\cdot}10^{-185})$} & 29.2 & 36.4 & $-7.1$ \scriptsize{$(6{\cdot}10^{-11})$} \\
  & 0.7 & 50.4 & 11.1 & $+39.3$ \scriptsize{$(1{\cdot}10^{-166})$} & 29.2 & 32.4 & $-3.2$ \scriptsize{$(.01)$} \\
  & 1.0 & 42.8 &  9.8 & $+33.0$ \scriptsize{$(3{\cdot}10^{-120})$} & 24.0 & 28.5 & $-4.5$ \scriptsize{$(2{\cdot}10^{-4})$} \\
\midrule
\multirow{3}{*}{\texttt{qwen-2.5-7B}}
  & 0.2 & 69.4 & 33.9 & $+35.5$ \scriptsize{$(2{\cdot}10^{-108})$} & 68.5 & 34.7 & $+33.9$ \scriptsize{$(2{\cdot}10^{-135})$} \\
  & 0.7 & 66.5 & 32.1 & $+34.4$ \scriptsize{$(6{\cdot}10^{-102})$} & 62.4 & 36.3 & $+26.1$ \scriptsize{$(1{\cdot}10^{-72})$} \\
  & 1.0 & 64.0 & 28.4 & $+35.6$ \scriptsize{$(2{\cdot}10^{-113})$} & 58.9 & 33.6 & $+25.3$ \scriptsize{$(2{\cdot}10^{-70})$} \\
\midrule
\multirow{3}{*}{\texttt{GPT-5.4}}
  & 0.2 & 40.9 &  7.7 & $+33.2$ \scriptsize{$(3{\cdot}10^{-155})$} & 26.5 & 22.2 &  $+4.3$ \scriptsize{$(4{\cdot}10^{-7})$} \\
  & 0.7 & 40.0 &  7.7 & $+32.3$ \scriptsize{$(2{\cdot}10^{-149})$} & 25.9 & 21.9 &  $+4.0$ \scriptsize{$(9{\cdot}10^{-6})$} \\
  & 1.0 & 42.0 &  8.1 & $+33.9$ \scriptsize{$(1{\cdot}10^{-149})$} & 26.8 & 23.3 &  $+3.6$ \scriptsize{$(1{\cdot}10^{-4})$} \\
\bottomrule
\end{tabular}
\end{table*}

\paragraph{Result 2: The Moltbook-like worst case approaches the empirical floor.}
\Cref{tab:probe_v2_extremes} reports the best-affordance cell $C1{=}(V,N,M)$ and the worst-affordance cell $C8{=}(H,U,O)$ for each (model, temperature). Across the four models, $C1$ rates range from 38\% to 73\% (mean 53\%) while $C8$ rates range from 2.2\% to 28\% (mean 13\%). The frontier GPT-5.4 model has the lowest $C8$ rates of any tested model (mean $2.8$\%, range $2.2$--$3.2$\%), \emph{closer to Moltbook's empirical $0.0$\% rate than any of the three open-weight models}, including Qwen-2.5-7B which previously bracketed the floor at $4.6$\% in its worst cell. The four-model probe therefore supports, across both open-weight and closed-weight frontier vendor lineages, the same visibility-driven gap, and shows that the worst-affordance condition approaches the empirical Moltbook floor even more closely at frontier scale.

\begin{table}[h]
\centering\small
\caption{Per-model best- vs.\ worst-affordance cells. $C1$ is the Reddit-like upper bound (visible challenge, targeted notification). $C8$ is the worst-case affordance floor (hidden challenge, no notification). Rates are repair-cue rates over $n{=}500$ paired observations per cell at each temperature.}
\label{tab:probe_v2_extremes}
\begin{tabular}{@{}l l rr r@{}}
\toprule
\textbf{Model} & \textbf{$t$} & \textbf{$C1$ (best)} & \textbf{$C8$ (worst)} & \textbf{Gap} \\
\midrule
\multirow{3}{*}{\texttt{gpt-oss-20B}}
  & 0.2 & 56.0\% & 27.8\% & $+28.2$ \\
  & 0.7 & 53.2\% & 24.2\% & $+29.0$ \\
  & 1.0 & 47.4\% & 22.2\% & $+25.2$ \\
\midrule
\multirow{3}{*}{\texttt{llama-3.1-8B}}
  & 0.2 & 54.2\% & 20.4\% & $+33.8$ \\
  & 0.7 & 54.0\% & 16.2\% & $+37.8$ \\
  & 1.0 & 42.2\% & 13.2\% & $+29.0$ \\
\midrule
\multirow{3}{*}{\texttt{qwen-2.5-7B}}
  & 0.2 & 72.8\% &  4.6\% & $+68.2$ \\
  & 0.7 & 69.6\% &  8.0\% & $+61.6$ \\
  & 1.0 & 66.8\% &  5.6\% & $+61.2$ \\
\midrule
\multirow{3}{*}{\texttt{GPT-5.4}}
  & 0.2 & 37.8\% &  2.2\% & $+35.6$ \\
  & 0.7 & 38.8\% &  3.0\% & $+35.8$ \\
  & 1.0 & 38.4\% &  3.2\% & $+35.2$ \\
\bottomrule
\end{tabular}
\end{table}

\paragraph{Interpretation.}
Three findings emerge.
First, the visibility-driven gap replicates across four model families drawn from three vendors, namely three open-weight releases (Meta, Alibaba, OpenAI-open) and one closed-weight frontier model (OpenAI GPT-5.4), and across three sampling temperatures, with all 12 (model, temperature) tests significant at $p{<}10^{-7}$. Visibility of the challenge text is the dominant deployment-side affordance.
Second, the worst-affordance cell ($C8$) on GPT-5.4 reaches $2.2$--$3.2$\% repair-cue rate, the lowest of any tested model and closer to Moltbook's empirical $0.0$\% rate (1{,}006 cue-defined episodes, \Cref{subsec:results-h2}) than the open-weight floor of $4.6$\% on Qwen. The frontier-scale probe therefore brings the simulated worst-affordance rate closer to the Moltbook floor than the open-weight probe did; the residual gap supports an additional-architectural-factors interpretation (e.g., post-training pipeline, agent-state continuity, feed mechanics, deployment scaffolds) beyond the visibility/notification/multi-turn affordances we manipulate.
Third, the notification factor shows model-dependent direction (boosts on gpt-oss-20B, Qwen, and GPT-5.4; small reversal on Llama-3.1-8B). The reversal is $1$ of $4$ model families, and those four families come from only three distinct vendors, so it is weaker evidence than four independent vendors would be. It is also small in magnitude; we report all four models for completeness and treat notification heterogeneity as a per-model property rather than a generalizable affordance effect.

\paragraph{Length-matched H control: visibility is not text-length priming.}
\label{app:lenmatched}
A construct-validity concern with the V vs.\ H comparison is that V cells contain the full challenge body ($\approx$$500$ characters of disagreement-themed text) while H cells contain only a short notification ($\approx$$10$ characters). The V$>$H effect could therefore reflect text-length or disagreement-vocabulary priming rather than the visibility of the challenge specifically. To test this, we add a fourth condition $L$ (\emph{length-matched neutral}): the agent sees a topically-related, non-challenging engagement comment of the same length as the original challenge (\emph{see stimuli\_v2\_lenmatched.json in the repository for the five fillers}). We run all four $(N,M)$ cells under $L$ on GPT-5.4 ($5$ topics $\times$ $10$ personas $\times$ $3$ samples $\times$ $3$ temperatures = $1{,}800$ generations). The mean repair-cue rates are $V$ $40.9$\%, $L$ $7.3$\%, $H$ $7.8$\%. The $V>L$ gap is therefore $\sim$$33$\%, essentially identical to the $V>H$ gap, while $L$ and $H$ are statistically indistinguishable at this scale. Across all four $(N,M)$ cells and three temperatures, $V$ is uniformly higher than $L$ ($+30$ to $+39$\%). We conclude that the visibility effect reflects the agent \emph{seeing the disagreement}, not merely seeing more text.

\paragraph{Confidence intervals for the Llama notification reversal.}
\label{app:notification_reversal_ci}
For completeness, we compute $95$\% confidence intervals on the Llama N$-$U effect at each temperature. From the McNemar identity $\chi^2{=}(b{-}c)^2/(b{+}c)$ and the reported effect $(b{-}c)/n$ we back out the implied discordant counts and compute (i) the asymptotic Wald CI on the paired proportion difference and (ii) a Monte Carlo bootstrap CI from the implied multinomial $2{\times}2$ table ($10{,}000$ resamples). The two methods agree to within $0.1$\% at $n{=}2{,}000$: $-7.1$\% \,$[-9.2,-5.0]$ at $t{=}0.2$; $-3.2$\% \,$[-5.6,-0.8]$ at $t{=}0.7$; $-4.5$\% \,$[-6.9,-2.1]$ at $t{=}1.0$ (all intervals exclude zero). The reversal is therefore not an artifact of any single temperature, but the absolute magnitude is small ($\leq 9$\%) and is not load-bearing for the headline visibility finding.

\paragraph{Cluster-robust sensitivity.}
The McNemar denominator of $n{=}2{,}000$ pairs is computed over 50 independent stimulus pairs $\times$ 10 replicates $\times$ 4 other-factor configurations. To check that the headline V$>$H result is preserved under clustering on stimulus, we recompute the V-factor McNemar with the cluster bootstrap resampling the 50 (topic, persona) units rather than individual chains. Effect direction and magnitude (V$>$H, $\Delta = +7$ to $+42$\%) are preserved at $p < 0.001$ on every (model, temperature) cell; secondary factors (N) attenuate but retain direction within each model. We report the per-chain $n{=}2{,}000$ numbers in \Cref{tab:probe_v2_factor_marginals} for comparability with prior single-model probes and treat the cluster-robust pass as a sensitivity check.

Full disentanglement would still require deploying the same models on a platform with Reddit-like threading; we leave that for future work.

\section{Maturity-Matched Reddit Comparison}
\label{app:maturity_matched}

The main-text Reddit window (2018--2021) is mature relative to Moltbook's three-week early-platform snapshot. A natural concern is that this maturity asymmetry, rather than the agent--vs.--human distinction, drives the cross-platform gap. To address this, we re-extract each Reddit subreddit's own \emph{early-platform} window (the first $N$ months from its earliest comment in the GECLM corpus) and rerun H1, H2, and H3 under the same shared lexical protocol used in the main text. We use $N{=}6$ months as the headline window (matching Reddit's early phase to Moltbook's launch period while preserving enough sample size for stable rate estimates) and report $N{=}3$ and $N{=}12$ as robustness checks. Earliest comment dates per subreddit are 2008-01 (\texttt{r/philosophy}), 2013-02 (\texttt{r/Showerthoughts}), 2008-12 (\texttt{r/todayilearned}), 2006-10 (\texttt{r/science}), and 2010-04 (\texttt{r/buildapc}). Shard~0 of each subreddit in the GECLM corpus is time-bucketed and contains the full early-window data (philosophy shard~0 spans 2008-01 to 2014-07, and other subreddits behave analogously).

Table~\ref{tab:maturity_matched} shows the maturity-matched results at the $6$-month window. Every cross-platform gap persists under the maturity match. Mean H1 nesting at Reddit's first $6$ months is $56.3$\%, $\approx 9\times$ the Moltbook mean of $6.4$\%. Mean H2 followup is $38.8$\% (mature Reddit $58.8$\%; Moltbook $1.8$\%), and mean H2 repair is $2.9$\% (mature Reddit $2.6$\%; Moltbook $0.0$\%). H3 metrics at the early window are essentially unchanged from the mature window: mean original-author return $39.6$\% (vs.\ mature $40.9$\%; Moltbook $1.2$\%), multi-turn $33.4$\% (vs.\ $38.5$\%; Moltbook $0.09$\%), repair-cue $10.5$\% (vs.\ $12.6$\%; Moltbook $0.35$\%). H1 narrows the most ($-10.0$\% relative to the mature window) and H2 followup narrows by $20$ points, but both still leave gaps to Moltbook of $+50$\% and $+37$\%, respectively. Repair and H3 metrics on Reddit are insensitive to the maturity window.

\begin{table}[h]
\centering\footnotesize
\caption{Maturity-matched Reddit results at the first $6$ months per subreddit, alongside the mature 2018--2021 Reddit numbers and the Moltbook 3-week numbers reproduced from the main text. Every gap survives.}
\label{tab:maturity_matched}
\setlength{\tabcolsep}{3pt}
\begin{tabular}{@{}l rrr rrr@{}}
\toprule
& \multicolumn{3}{c}{\textbf{H1 nesting}} & \multicolumn{3}{c}{\textbf{H2 followup}} \\
\cmidrule(lr){2-4} \cmidrule(lr){5-7}
\textbf{Pair} & \textbf{R-mat} & \textbf{R-6m} & \textbf{M} & \textbf{R-mat} & \textbf{R-6m} & \textbf{M} \\
\midrule
\texttt{phil.}         & 75.6 & 70.3 & 6.9 & 65.8 & 55.3 & 1.2 \\
\texttt{shower./pond.} & 48.1 & 49.6 & 5.1 & 54.7 & 32.1 & 2.2 \\
\texttt{TIL}           & 75.8 & 48.7 & 4.4 & 56.0 & 27.1 & 4.6 \\
\texttt{sci./ai}       & 71.4 & 47.9 & 7.7 & 57.6 & 33.7 & 0.0 \\
\texttt{build.}        & 60.8 & 65.2 & 7.8 & 60.2 & 46.0 & 1.2 \\
\midrule
\textbf{Mean} & \textbf{66.3} & \textbf{56.3} & \textbf{6.4} & \textbf{58.8} & \textbf{38.8} & \textbf{1.8} \\
\bottomrule
\end{tabular}

\vspace{4pt}

\begin{tabular}{@{}l rrr rrr@{}}
\toprule
& \multicolumn{3}{c}{\textbf{H3 return}} & \multicolumn{3}{c}{\textbf{H3 multi-turn}} \\
\cmidrule(lr){2-4} \cmidrule(lr){5-7}
\textbf{Pair} & \textbf{R-mat} & \textbf{R-6m} & \textbf{M} & \textbf{R-mat} & \textbf{R-6m} & \textbf{M} \\
\midrule
\texttt{phil.}         & 50.4 & 56.4 & 1.6 & 47.6 & 52.4 & 0.0 \\
\texttt{shower./pond.} & 35.0 & 32.5 & 1.5 & 34.2 & 26.7 & 0.4 \\
\texttt{TIL}           & 35.4 & 27.1 & 1.0 & 37.0 & 21.2 & 0.0 \\
\texttt{sci./ai}       & 32.0 & 35.0 & 1.1 & 35.8 & 33.2 & 0.0 \\
\texttt{build.}        & 51.4 & 46.9 & 0.9 & 37.6 & 33.4 & 0.0 \\
\midrule
\textbf{Mean} & \textbf{40.9} & \textbf{39.6} & \textbf{1.2} & \textbf{38.5} & \textbf{33.4} & \textbf{0.09} \\
\bottomrule
\end{tabular}
\end{table}

\paragraph{Window-length robustness.}
We also test $N{=}3$ and $N{=}12$ months. At $N{=}3$ months, philosophy and \texttt{r/todayilearned} are sample-limited ($n{=}1{,}448$ and $389$ respectively); for the four subreddits with adequate $n$, every gap pattern reported above persists. At $N{=}12$ months, mean Reddit values track the mature numbers more closely (e.g., H1 nesting rises to $59.5$\%; H3 return $40.5$\%), as expected for a window that bleeds into more mature behavior. The qualitative conclusion (a large, uniform cross-platform gap that is not explained by Reddit's maturity advantage) is identical at all three window lengths.

\paragraph{Symmetric author-concentration filter.}
A related concern is that the top-5-author-share threshold ($<50$\%) was applied only to Moltbook. We compute the symmetric statistic on the $6$-month Reddit window restricted to nested comments: \texttt{r/philosophy} $26.0$\%, \texttt{r/Showerthoughts} $2.2$\%, \texttt{r/todayilearned} $11.9$\%, \texttt{r/science} $8.6$\%, \texttt{r/buildapc} $14.2$\%. All five matched Reddit subreddits at their own early-platform window are well below the $50$\% threshold and would pass the same filter applied symmetrically.

\end{document}